\documentclass[letterpaper, reprint, showkeys, footinbib, prl, aps, superscriptaddress, longbibliography, nobalancelastpage]{revtex4-1}

\usepackage[pdftex, breaklinks=true]{hyperref}
\hypersetup{
    pdffitwindow=true,     
    pdfstartview={FitH},    
    pdfnewwindow=true,      
    colorlinks=true,       
    linkcolor=red,      
    citecolor=blue,        
    urlcolor=blue           
}
\usepackage[utf8]{inputenc}
\DeclareSymbolFont{slenderlargesymbols}{OMX}{ccex}{m}{n}
\DeclareMathSymbol{\prod}{\mathop}{slenderlargesymbols}{"51}
\usepackage{amsfonts, amssymb, amsmath}
\usepackage{amsthm}
\usepackage{mathtools}
\usepackage{braket} 
\usepackage{multirow}
\usepackage[T1]{fontenc}

\newtheoremstyle{cited}%
  {3pt}
  {3pt}
  {\itshape}
  {}
  {\bfseries}
  {.}
  {.5em}
  {\thmname{#1} \thmnumber{#2} \thmnote{\normalfont#3}}
\theoremstyle{cited}
\newtheorem{citedprop}{Proposition}

\renewcommand\bra[1]{{\langle{#1}|}}
\makeatletter
\renewcommand\ket[1]{%
  \@ifnextchar\bra{\k@t{#1}\!}{\k@t{#1}}%
}
\newcommand\k@t[1]{{|{#1}\rangle}}
\makeatother
\DeclarePairedDelimiter{\abs}{\lvert}{\rvert}
\DeclarePairedDelimiter{\norm}{\lVert}{\rVert}
\DeclareMathOperator{\tr}{\text{tr}}

\usepackage{xcolor}
\newcommand{\correct}[1]{{\color{black} #1}}
\usepackage[normalem]{ulem}

\newcommand{\bw}[1]{{\color{black} #1}}
\newcommand{\bww}[1]{{\color{black} #1}}
\newcommand{\hl}[1]{{\color{black} #1}}

\newtheorem{theorem}{Theorem}

\newtheorem{definition}{Definition}
\newtheorem{coro}{Corollary}
\newtheorem{lemma}{Lemma}

\usepackage{tikz}
\usetikzlibrary{decorations.pathreplacing}
\makeatletter
\def\underbrace#1{%
   \@ifnextchar_{\tikz@@underbrace{#1}}{\tikz@@underbrace{#1}_{}}}
\def\tikz@@underbrace#1_#2{%
   \tikz[baseline=(a.base)] {\node[inner sep=2] (a) {\(#1\)};
   \draw[line cap=round,decorate,decoration={brace,amplitude=4pt}]
     (a.south east) -- node[pos=0.5,below,inner sep=7pt] {\(\scriptstyle #2\)} (a.south west);}}
 
\def\overbrace#1{%
   \@ifnextchar^{\tikz@@overbrace{#1}}{\tikz@@overbrace{#1}^{}}}
\def\tikz@@overbrace#1^#2{%
   \tikz[baseline=(a.base)] {\node[inner sep=2] (a) {\(#1\)};
   \draw[line cap=round,decorate,decoration={brace,amplitude=4pt}]
     (a.north west) -- node[pos=0.5,above,inner sep=7pt] {\(\scriptstyle #2\)} (a.north east);}}
\makeatother

\begin{document}
\title{\correct{Robust in Practice: Adversarial Attacks on Quantum Machine Learning}}

\author{Haoran Liao}
\email[E-mail: ]{haoran.liao@berkeley.edu}
\affiliation{Department of Physics, University of California, Berkeley, CA 94720, USA}
\affiliation{Berkeley Quantum Information and Computation Center, University of California, Berkeley, CA 94720, USA}

\author{Ian Convy}
\affiliation{Department of Chemistry, University of California, Berkeley, CA 94720, USA}
\affiliation{Berkeley Quantum Information and Computation Center, University of California, Berkeley, CA 94720, USA}

\author{William J. Huggins}
\affiliation{Department of Chemistry, University of California, Berkeley, CA 94720, USA}
\affiliation{Berkeley Quantum Information and Computation Center, University of California, Berkeley, CA 94720, USA}

\author{K. Birgitta Whaley}
\affiliation{Department of Chemistry, University of California, Berkeley, CA 94720, USA}
\affiliation{Berkeley Quantum Information and Computation Center, University of California, Berkeley, CA 94720, USA}

\date{\today}

\begin{abstract}
\correct{State-of-the-art classical neural networks are observed to be vulnerable to small crafted adversarial perturbations. A more severe vulnerability has been noted for quantum machine learning (QML) models classifying Haar-random pure states.} This stems from the concentration of measure phenomenon, a property of the metric space when sampled probabilistically, and is independent of the classification protocol. \correct{In order to provide insights into the adversarial robustness of a quantum classifier on real-world classification tasks, we focus on the adversarial robustness in classifying a subset of encoded states that are smoothly generated from a Gaussian latent space. We show that the vulnerability of this task is considerably weaker than that of classifying Haar-random pure states.} \correct{In particular, we find only mildly polynomially decreasing robustness in the number of qubits, in contrast to the exponentially decreasing robustness when classifying Haar-random pure states and} \bww{suggesting that QML models can be useful for real-world classification tasks.}

\end{abstract}

\maketitle

\section{Introduction}

Quantum machine learning (QML) protocols, by exploiting quantum mechanics principles, such as superposition, tunneling, and entanglement \cite{biamonte_quantum_2017}, have given hope of outperforming their classical counterparts, even with noisy intermediate-scale quantum (NISQ) \cite{preskill_quantum_2018} hardware in the near-term \cite{xia_quantum-enhanced_2020}. For classification tasks where statistical patterns can be revealed in complex feature spaces, the high-dimensional Hilbert space of sizable quantum systems offers a naturally advantageous starting ground for QML models. However, many state-of-the-art classical machine learning models, such as deep neural networks with complicated internal feature mappings, have been shown vulnerable to small crafted perturbations to the input, namely adversarial examples \correct{\cite{Szegedy2014, Goodfellow2014}}. These are intentional worst-case perturbations to the original samples with an imperceptible difference that are nevertheless misclassified by the classifier.  This not only raises questions as to why well-performing classifiers suffer from such instabilities but also poses security threats to machine learning applications that emphasize reliability, such as in spam filtering \cite{dalvi_adversarial_2004}. To understand this unreliable behavior, the transferability of these attacks across different architecture and the robustness against perturbations has led to extensive investigations in the classical machine learning community in recent years \cite{Chakraborty2018,biggio_wild_2018,huang_adversarial_2011}. Notably, some geometric and probabilistic arguments, based on curvatures of decision boundaries \cite{Fawzi} and the concentration of measure \cite{Mahloujifar2019, Gilmer, Diochnos2018,Fawzia,Fawzi2018}, have been employed to quantify the risk of adversarial attacks in various settings. \correct{It has been shown that any classifier will have an adversarial robustness that is increasingly reduced by the dimension of the space on which it classifies, given the concentration of measure phenomenon in certain metric probability spaces} \cite{Mahloujifar2019}. This has raised attention in the QML community where the models take advantage of the high dimensionality of quantum systems \cite{Liu, Lu, Du2020, Guan_robustness_2020} 

\correct{The concentration of measure is a phenomenon that describes the fact that, in certain metric probability spaces, points tend to gather around the boundaries of subsets having at least one half of the probability measure. As a result, there is generically a high probability of obtaining values close to the average for any reasonably smooth function that is evaluated on the distribution \cite{ledoux_concentration_2001,vitali_d_milman_gideon_schechtman_asymptotic_2002,mcclean_barren_2018, popescu_entanglement_2006, muller_concentration_2011}. This means \bw{that} when samples are selected from such a concentrated space, \bw{the confidences predicted by the classifier tends to accumulate \bww{around}} the critical value separating the correct and incorrect classes.} As such, small \bw{targeted} perturbations can \bw{then easily} move the samples across the decision boundary. In particular, \bw{it has been recognized that} this phenomenon can lead to extreme vulnerabilities of any quantum classifier on high-dimensional Haar-random pure states\bw{~\cite{Liu}}. Nevertheless, there is no indication of whether such vulnerability exists when classifying on a subset of encoded pure states \correct{in a realistic task,} \bw{such as using a quantum classifier on classical images encoded in pure states}.

In this paper, \correct{we approach the task of classifying quantum states from a geometric perspective}. \correct{The quantum classifier divides the Hilbert space into subsets, each of which belongs to a certain class.} We use this perspective \bw{here} to study aspects of the problem that are relevant to practical applications of QML. \correct{In a practical classification task, such as in recognizing natural images, the samples \bw{to be classified} can be generated from a Gaussian latent space by \bw{one of a number of  commonly-used} generative models \cite{Rezende,kingma_auto-encoding_2014,dinh_density_2017,Goodfellow,bojanowski_optimizing_2018,arjovsky_wasserstein_2017}. The success of these models for real-world data generation \bw{ensures} that the focus on QML models classifying a subset of encoded pure states, where these states are sampled from a distribution that is \correct{\textit{smoothly}} mapped from a Gaussian latent space \cite{Fawzi2018}, \bw{will yield insight into the vulnerability of QML models in a real-world classification task. This contrasts with the previous analysis of the vulnerabilities when classifying Haar-random pure states \cite{Liu}.}}

We demonstrate that the \correct{\textit{adversarial robustness}} \hl{over this generated distribution} decreases as $\mathcal{O}(1/\sqrt{n})$ in the number of qubits $n$, with the scaling measured in the trace norm. This decline in the robustness is mild, indicating a quantum classifier can be robust to attacks on high dimensional quantum states. In contrast, when considering \correct{\textit{prediction-change}} adversarial settings where the inputs are pure states drawn Haar-randomly, we show that the robustness decreases as $\mathcal{O}(1/2^n)$ in the number of qubits \correct{$n$}, implying extreme vulnerabilities to attacks in high-dimensional quantum systems. \correct{This second case parallels the result of reference \cite{Liu}, which considered \textit{error-region} adversarial settings and found the robustness
\bw{also} decreases as $\mathcal{O}(1/2^n)$ here.} \correct{However, we argue that \bw{the} extreme vulnerability \bw{in this setting} is \bw{not of concern in practice, since the states to be classified are always} sampled from a distribution over some subsets of states, rather than \bw{from} the Haar-random distribution over the entire set of pure states.}

The \bw{rest of the} paper is structured as follows. In Section~\ref{sec:background}, we introduce the set-ups and preliminaries in both classical and quantum adversarial attacks. In Section~\ref{sec:problems_practice}, 
\bww{we describe the prediction-change adversarial setting, which is often more relevant to real-world classification tasks \hl{than the} previously employed error-region adversarial setting. We then derive} the prediction-change adversarial robustness of any quantum classifier on Haar-randomly \hl{distributed} pure states and explain its practical limitations. In Section~\ref{sec:incorporation_generative}, we derive the main results on the adversarial robustness of any quantum classifier classifying a smoothly generated distribution over a subset of encoded pure states of interest, and propose a feasible modification to any quantum classifier to lower bound unconstrained adversarial robustness. In Section~\ref{sec:comparison}, a summary and discussion of the derived robustness \hl{over the two types of distribution} are presented.

\section{Background} \label{sec:background}
\subsection{Classical Adversarial Attacks}
Classical adversarial attacks were introduced to analyze the instability of deep neural networks caused by a small change to the input sample. Classically, \bw{the} confidence is often quantified as the probability corresponding to the label class in the output normalized discrete distribution, e.g., the largest softmax value in the output vector in a multi-class logistic-regression convolutional neural network. As numerically shown in various works, such an attack results in a significant drop in the confidence in the correct class \cite{Szegedy2014,Kurakin2019a,Kurakin2019,biggio_wild_2018}, and may also bring \correct{a} significant increase in the confidence in the incorrect class \cite{Goodfellow2014}.
So far, some arguments have been proposed to explain the vulnerabilities of various classifiers to adversarial attacks and their transferability \cite{Goodfellow2014,Bubeck2018,Charles,engstrom_adversarial_2019,Fawzia}, yet no conclusive consensus has been established \cite{Gopfert2019}. 

\par The most common type of adversarial attack is the evasion attack where the adversary does not interfere with the training phase of a classifier and perturbs only the testing samples \cite{Chakraborty2018}. The adversary can devise white-box attacks if it possesses total knowledge about the classifier architecture, or otherwise, it can devise black-box attacks relying on the transferability \cite{Chakraborty2018, biggio_wild_2018}. We shall focus on white-box evasion attacks.
\par We introduce some notations and definitions used in this paper. Let $(\mathcal{X}, \mathrm{d}, \mu)$ denote the sample set $\mathcal{X}$ with a metric $\mathrm{d}$ and a probability measure $\mu$. The notation $x \leftarrow \mu$ denotes that a sample $x$ is drawn with a probability measure $\mu$. $\mathcal{L}$ denotes the countable label set. For a subset $\mathcal{S} \subseteq \mathcal{X}$, we let $\mathrm{d}(x, \mathcal{S}) = \inf\{\mathrm{d}(x, y) | y \in \mathcal{S}\}$ and let $B_\epsilon(x) = \{x' | \mathrm{d}(x, x') \leq \epsilon\}$ be the $\epsilon$-neighborhood of $x$, where $\mathrm{d}$ is the metric on $\mathcal{X}$. We also let $\mathcal{S}_\epsilon=\{x | \mathrm{d}(x, \mathcal{S}) \leq \epsilon\}$ be the $\epsilon$-expansion of $\mathcal{S}$. $h$ is a hypothesis or a trained classifier that maps each $x\in\mathcal{X}$ to a predicted label $l \in \mathcal{L}$. $c$ is the ground-truth function that maps each $x\in \mathcal{X}$ to a correct label $l \in \mathcal{L}$. $h^l$ denotes the set of samples classified as label $l$, namely $h^l = \{x \in \mathcal{X} | h(x) = l\}$. The error region $\mathcal{M}$ is the set of samples on which the hypothesis disagrees with the ground-truth, namely $\mathcal{M}=\{x| h(x) \neq c(x)\}$. We define the risk as $R(h,c) = \Pr_{x\leftarrow \mu}[h(x) \neq c(x)]=\mu(\mathcal{M})$. 

The two relevant types of evasion attacks studied here are based on the error region and the prediction change. In an error-region attack, the ground-truth function $c$ is accessible and an attack occurs when a perturbation in the sample causes $h$ to disagree with $c$. In contrast, a prediction-change attack emphasizes the instability of $h$: an attack occurs when a perturbation results in a different prediction by $h$, and $c$ is irrelevant. The precise definitions of these two types of attacks are as follows.

\begin{definition} \label{def:er_risk}
The error-region adversarial risk under $\epsilon$-perturbation is the probability of drawing a sample such that its $\epsilon$-neighborhood intersects with the error region,
\begin{equation*} 
R^{ER}_{\epsilon}(h, c, \mu) = \Pr_{x\leftarrow \mu} [\exists x' \in B_\epsilon(x) | h(x') \neq c(x')].
\end{equation*}
\end{definition}

\begin{definition} \label{def:pc_risk_1}
The prediction-change adversarial risk under $\epsilon$-perturbation is the probability of drawing a sample such that its $\epsilon$-neighborhood contains a sample with a different label,
\begin{equation*} 
R^{PC}_{\epsilon}(h, \mu)=\Pr_{x \leftarrow \mu}[\exists x' \in B_\epsilon(x) | h(x) \neq h(x')],
\end{equation*}
equivalently,
\begin{equation*}
R^{PC}_{\epsilon}(h, \mu)=\Pr_{x\leftarrow\mu}\left[\min_{x'\in \mathcal{X}}\{\mathrm{d}(x',x)|h(x') \neq h(x)\} \leq \epsilon\right].
\end{equation*}
\end{definition}

 In either type of attack, we call the nearest misclassified examples as the adversarial examples. We say that $h$ is more robust if the induced risk of either type is lower for a certain $\epsilon$-perturbation. We shall refer to the minimal $\epsilon$-perturbation to $x$ resulting in an adversarial example as the adversarial perturbation or the robustness of $x$ with $h$. In contrast, we shall \correct{quantify} the adversarial robustness of $h$ as the size of $\epsilon$ necessary for the adversarial risk of $h$ to be upper bounded by some constant. \correct{The main result of this paper is an upper bound on the adversarial robustness of any quantum classifier when the input states are smoothly generated from a Gaussian latent space.}

\subsection{Quantum Adversarial Attacks}\label{sec:qml_attack_setup}
For our work, a quantum classifier is a quantum channel $\mathcal{E}$ that assigns labels \bww{$l$} with some set of positive-operator-valued measures (POVMs) $\{\hl{\Pi_l}\}$. The quantum classifier takes in an ensemble of
identically prepared copies of a state and assigns the state a label \bww{$l$}. The confidence of a prediction is quantified as the expectation value of the POVM for the prediction \bww{$l$}, namely $\tr(\mathcal{E}(\rho)\hl{\Pi_l})$ for an input density matrix $\rho$. \correct{We do not consider the number of copies of a state that is required to implement any specific quantum classification protocol.} To measure the perturbation size, the natural choice of metric on quantum states -- the trace distance -- can be shown to generate an upper bound on the difference between their quantum classification confidence (see Appendix~\ref{append:metric}), which implies that no small variation can induce a large swing in the predictive confidence. This property of the trace distance is a consequence of its interpretation as the achievable upper bound on the total variation distance \footnote{Informally, total variation distance is the largest possible difference between the probabilities that the two distributions can assign to the same event.} between probability distributions arising from measurements performed on those quantum states \cite{nielsen_quantum_2010}. Furthermore, we show in Appendix~\ref{append:metric} that the Hilbert-Schmidt norm, the Bures distance, and the Hellinger distance between two quantum states all generate a similar upper bound. As a result, in quantum adversarial attacks, the adversary either perturbs the states near the decision boundary minimally to seek misclassification, or aims to maximize confidence change to any state with associated perturbations that are upper bounded by some considerable size in these norms, \correct{as illustrated in Figure~\ref{fig:1}}.  Our work analyzes primarily the risks due to the former objective. In Appendix~\ref{append:adv_TTN}, we also propose a method for the latter objective exploiting the reversibility of parametrized quantum circuits (see e.g. \cite{Huggins2019, benedetti_parameterized_2019}). We note that the latter adversarial setting is justified, since in order to assess the security of a classifier under attack, it is reasonable -- given a feasible space of modifications to the input data -- to assume that the adversary aims to maximize the classifier’s confidence in wrong predictions, rather than merely perturbing minimally \correct{in size} \cite{biggio_wild_2018}.

\begin{figure}[hbt!]
 \centering
 \includegraphics[width=0.55\linewidth]{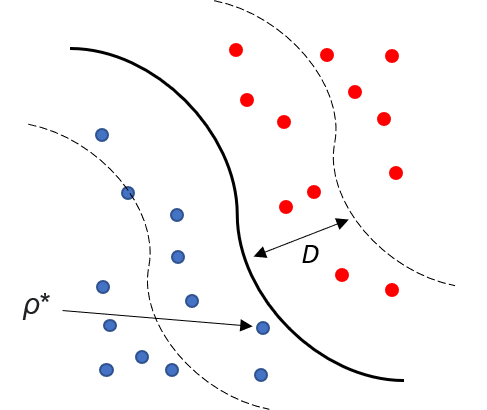}
 \caption{\correct{The solid curve depicts the decision boundary of a quantum classifier. The states in blue are classified in a different class \bw{from} the states in red. The metric is the trace distance. The trace distance between any pair of states generates an upper bound on the difference between their quantum classification confidences. \bw{Thus} $\rho^*$, the state closest to the decision boundary, is the ideal target of a prediction-change adversarial attack if the adversary aims to achieve misclassifications with minimal perturbations. On the other hand, if the adversary aims to maximize confidence change to any state with associated perturbations of size \bww{up to} $D$, then all states between the dashed lines can be perturbed to be misclassified, \bw{while} \bww{all} other states can be perturbed to get closer to the boundary, resulting in \bw{overall} decreased confidence in predicting the correct class. The concentration of measure phenomena implies that \bw{for a sufficiently large class, }samples tend to lie near the decision boundary.
 }}
 \label{fig:1}
\end{figure}

There are two natural set-ups of adversarial attacks in QML that can be specified. The first is when the input data to the classifier is already quantized and any data transmitted through the quantum communication network comes from an untrusted party. \correct{In this case, the adversary, \bw{who may} be the sender or an interceptor, \bw{can} perform an attack either by perturbing each of the transmitted density matrices, or by intercepting a fraction of the copies of the state and substituting them entirely} (see Appendix~\ref{append:metric}). In a broader context, our analysis can be extended to \correct{include} the instability of classifying quantum states subject to decoherence. We focus on this first set-up in the \bw{current} paper. The second set-up is when the input to the quantum classifier is classical. The quantum classifier encodes the classical data before classifying. Since the adversary is perturbing the classical input data, it is effectively attacking classically. \bw{If one views} such a quantum classifier as a black-boxed hypothesis function \correct{that maps each input to a class}, any classifier-agnostic classical \correct{analysis of adversarial robustness} can \bw{then} be directly applied. \correct{For example, reference \cite{Fawzi} analyzes the robustness of any classifier against random or semi-random perturbations, provided the curvature of the decision boundary is sufficiently small, \bw{while} reference \cite{Fawzi2018} analyzes the adversarial robustness of any classifier when classical input vectors are smoothly mapped from a Gaussian latent representation.}  

\subsection{Quantum Data Encoding}\label{sec:encoding}
\par \correct{We now explain the feature maps used throughout the paper.} Considering a normalized positive vector $u$ of length $n$, without loss of generality, we intuitively refer to it as a gray-scale image with $n$ pixels in this paper. We focus on a particular set of encoding schemes where \bww{the normalized gray-scale value of each pixel, i.e., $u_i\in[0,1], i = 1, \dots, n$,} is featurized into a qubit-encoding state $\ket{\phi_i}$. \bww{The product state $\ket{\phi}$ to be classified is a tensor product state of these qubit-encoded pixels} in the $2^n$-dimensional Hilbert space \cite{Stoudenmire2016,grant_hierarchical_2018,cao_cost-function_2020,martyn_entanglement_2020}, 
namely
\begin{equation} \label{eqn: product_states}
\begin{split}
\ket{\phi}&=\bigotimes_{i=1}^n\ket{\phi_i}
=\bigotimes_{i=1}^n\left[\cos\left(\frac{\pi}{2}u_i\right)\ket{0}+ \sin\left(\frac{\pi}{2}u_i\right)\ket{1}\right].
\end{split}
\end{equation}
 The qubit-encoding states, Eq.~\eqref{eqn: product_states}, do not require a quantum random access memory (QRAM) \cite{giovannetti_quantum_2008} and are efficient in time to
 prepare. Other schemes including amplitude encoding (see e.g., \cite{larose_robust_2020}) are not considered here. We note that some of our results are general and independent of the encoding scheme. We further generalize \correct{Eq.~\eqref{eqn: product_states}} to qudits. 
 \bww{In this case
 each pixel is mapped to a Hilbert space of higher dimension $d\geq2$, with} the \correct{coefficient of the} $j$-th component of the \correct{$i$-th} qudit state \bww{given by}
\begin{equation}\label{eqn:qudit}
    \ket{\phi_i}_j=\sqrt{\binom{d-1}{j-1}}\cos^{d-j}\left(\frac{\pi}{2}u_i\right)\sin^{j-1}\left(\frac{\pi}{2} u_i\right).
\end{equation}
These qudit states are special cases of what are known as spin-coherent states \cite{Stoudenmire2016}, and the qubit states in Eq.~\eqref{eqn: product_states} correspond to  $d=2$.

\subsection{Concentration of Measure Phenomenon}\label{sec:concentration_of_measure}
\par \correct{To describe this phenomenon,} let $\Sigma\subseteq \mathcal{X}$ be a Borel set \footnote{Borel sets are sets that can be constructed from open or closed sets through countable union, countable intersection, and relative complement}. The concentration function, defined as 
\begin{equation}\label{eqn:concentration_func}
    \alpha(\epsilon)=1-\inf_{\Sigma\subseteq\mathcal{X}}\left\{\mu(\Sigma_\epsilon) \big| \mu(\Sigma) \geq \frac{1}{2}\right\},
\end{equation} 
has a smaller value when more points are aggregated in the $\epsilon$-expansion of a sufficiently large \bww{set} $\Sigma$, for a fixed $\epsilon$. Informally, a space $\mathcal{X}$ exhibits a concentration of measure if $\alpha(\epsilon)$ decays very fast as $\epsilon$ grows, and we shall refer to it as a concentrated space. This is true for a simple example -- the standard Gaussian distribution $(\mathbb{R},\ell^2, \mathcal{N}(0,1))$. Looking at the Borel set $\Sigma=(-\infty,0)$ whose probability measure is $1/2$, the cumulative density outside its $\epsilon$-expansion, namely $\mathbb{R}\setminus\Sigma_\epsilon=(\epsilon,+\infty)$, decreases at least as fast as $\exp(-\epsilon^2/2)$ by the tail bound \cite{vershynin_four_2017}. One can invoke isoperimetric inequality \cite{borell_brunn-minkowski_1975} to show that this clustering occurs around any Borel set with measure at least $1/2$ and applies to any canonical $m$-dimensional Gaussian measure in the Euclidean space (see Appendix~\ref{append:proof_major_thm}). More formally, a family of $N$-dimensional spaces with corresponding concentration functions $\alpha_N(\cdot)$ is called a ($k_1$, $k_2$)-normal L\'evy family if $\alpha_{N}(\epsilon)\leq k_1 \exp(- k_2^2 \epsilon^2 N)$, where $k_1$ and $k_2$ are particular constants. Consequently, the measure is more concentrated for a higher dimension. Two notable normal L\'evy families are $\mathbb{SU}(N)$ and $\mathbb{SO}(N)$, both of which are equipped with the Hilbert-Schmidt norm $L^2$ and the Haar probability measure $\nu$ \cite{gromov_topological_1983,giordano_extremely_2007}. An implication of this phenomenon is that when points $x$ are drawn from a highly concentrated space, for any function $f$ varying not rapidly, we have $f(x)\approx\langle f\rangle$ with high probability. L\'evy's Lemma \cite{ledoux_concentration_2001,vitali_d_milman_gideon_schechtman_asymptotic_2002} constitutes a specific example of this.

\subsection{Related Work}
The work in \cite{Mahloujifar2019} considered any normal L\'evy family and derived the robustness for error-region adversarial attacks. The results show that for a nice classification problem \footnote{The precise definition of a nice classification problem can be found in Definition 2.3 in \cite{Mahloujifar2019}.}, if $\mu(\mathcal{M})=\Omega(1)$, the size of perturbations must be $\mathcal{O}(1/\sqrt{N})$ in order to have the error-region adversarial risk upper bounded by some constant, where $N$ is the dimension of the concentrated space. References \cite{Diochnos2018, Gilmer} studied some specific concentrated spaces and revealed the same scaling.

Reference \cite{Liu} transforms the classification of pure states $\ket{\phi}$ into that of unitaries $U$ in $\ket{\phi}=U\ket{\vec{0}}$ for some fixed initial state $\ket{\vec{0}}$. These quantum classifiers then classify samples drawn from $\mathbb{SU}(N)$ equipped with the Haar probability measure $\nu$ and the Hilbert-Schmidt norm, which is a $(\sqrt{2}, 1/4)$-normal L\'evy family. Therefore, if $\mu(\mathcal{M})>0$, the necessary condition on the perturbation size for the error-region adversarial risk to be bounded above by $1-\gamma$ for some $\gamma\in[0,1]$ is $\mathcal{O}(1/\sqrt{N})$. Precisely, to have $R^{ER}_{\epsilon}(h, c, \nu)\leq 1-\gamma$, the $\epsilon$-perturbation to any unitary must be upper bounded as \footnote{A concise proof of Eq.~\eqref{eqn:vul_epsilon} can be found in Appendix~\ref{append:nana_correction}.}
\begin{equation}\label{eqn:vul_epsilon}
    \epsilon\leq\sqrt{\frac{4}{N}}\left[\sqrt{\ln\left(\frac{\sqrt{2}}{\mu(\mathcal{M})}\right)}+\sqrt{\ln\left(\frac{\sqrt{2}}{\gamma}\right)}\right].
\end{equation}

\section{Problems with Practical Classifications} \label{sec:problems_practice}
The result in Eq.~\eqref{eqn:vul_epsilon} claims that when classifying unitaries in $\mathbb{SU}(N)$ with the Haar measure, given that an adversary can devise white-box attacks and $\mu(\mathcal{M})$ not exponentially suppressed by $N$, the robustness of any quantum classifier decreases polynomially in the dimension of the input $N$. This is daunting since the input has a dimension $N=d^n$ exponential in the number of qudits.

 To apply any result related to Eq.~\eqref{eqn:vul_epsilon}, a ground-truth function $c$ on $\mathbb{SU}(N)$ is needed to obtain the risk $\mu(\mathcal{M})$. However, $c$ may not be easily defined in a real-world machine learning task. For instance, it is challenging to define what
 constitutes a mistake for visual object recognition. After adding a perturbation to an image, it likely no longer corresponds to a photograph of a real physical scene \cite{elsayed_adversarial_2018}. Furthermore, it is difficult to define the labels for images undergoing gradual semantic change. All of these factors complicate the evaluation of $\mu(\mathcal{M})$. It thus motivates us to focus on prediction-change adversarial risks (see e.g., \cite{Diochnos2018,elsayed_adversarial_2018,Fawzi}) in order to avoid requiring access to the ground-truth. The following theorem and corollary then apply.

\par \begin{theorem} \label{thm: pc_risk_so}
Let $\mathbb{SU}(N)$ be equipped with the Haar measure $\nu$ and the Hilbert-Schmidt norm $L^2$. For any hypothesis $h: \mathbb{SU}(N)\rightarrow \mathcal{L}$ that is not a constant function, let $\eta \in [0, 1/2]$ determine the measure of the dominated class such that $\nu(h^l) \leq 1-\eta, \forall l \in \mathcal{L}$. Suppose $U \in h^l$, $V \notin h^l$ and a perturbation $U \rightarrow V$ occurs, where $\norm{U-V}_2 \leq \epsilon$. If the prediction-change adversarial risk $R^{PC}_\epsilon(h, \nu)\leq1-\gamma$, then $\epsilon$ must satisfy
\begin{equation} \label{eqn: pc_risk_so}
    \epsilon \leq \sqrt{ \frac{4}{N}}\left[\sqrt{\ln{\left(\frac{2\sqrt{2}}{\eta} \right)}}+\sqrt{\ln{\left(\frac{2\sqrt{2}}{\gamma} \right)}}\right]. 
\end{equation}
\end{theorem}
\correct{
\bw{It is evident from Eq.~\eqref{eqn: pc_risk_so} that} the upper bound on the size of the perturbation $\epsilon$ is suppressed \bww{as} the dimension $N$ of the space increases. It is also suppressed when the measure of the \bw{dominated} class ($1-\eta$) decreases and when the tolerance on the adversarial risk ($1-\gamma$) decreases.}

\begin{coro} \label{cor:to_thm_pc_risk}
 With $\rho=U\ket{\vec{0}}\bra{\vec{0}}U^\dagger$ and $\sigma=V\ket{\vec{0}}\bra{\vec{0}}V^\dagger$, Eq.~\eqref{eqn: pc_risk_so} translates to a necessary upper bound in the trace norm between the pure-state density matrices
 \begin{equation*}
     \norm{\rho-\sigma}_1\leq \frac{4}{N}\lambda_1=\Omega(d^{-n}).
 \end{equation*}
 With the qudit encoding in Eq.~\eqref{eqn:qudit}, a naive translation of this necessary upper bound to that in the $\ell^1$ norm of the encoding vectors $u$ and $v$ gives,
  \begin{equation*}
     \norm{u-v}_1\leq\frac{2n}{\pi}\cos^{-1}\left[\left(1-\frac{2}{N}\lambda_1\right)^{\frac{1}{(d-1)n}}\right]=\Omega(d^{-\frac{n}{2}}\sqrt{n}),
 \end{equation*}
where $N=d^n$ and $\lambda_1=[\ln(2\sqrt{2}/\eta)]^{1/2}+[\ln(2\sqrt{2}/\gamma)]^{1/2}$ with $\eta$ and $\gamma$ defined in Theorem~\ref{thm: pc_risk_so}.
\end{coro}
\noindent The proofs can be found in Appendix~\ref{append:proof_thm_pc_risk_su} and~\ref{append:proof_cor}. The interpretation of Theorem~\ref{thm: pc_risk_so} and Corollary~\ref{cor:to_thm_pc_risk} is clear: given that no class occupies Haar-measure $1$, any quantum classifier on quantum states is more vulnerable to prediction-change adversarial attacks on higher-dimensional pure states drawn Haar-randomly, with the robustness decaying exponentially in the number of qudits.

In what follows, we apply this theorem to a practical task by presenting two perspectives on the application, in order to illustrate the limitations of the theorem. Suppose that the objective of the practical task is to classify a subset of quantum states, for example, the pure product states in Section~\ref{sec:encoding} that encode images displaying a digit $0$ or $1$. On one hand, if we label unitaries not related to an actual image, together with unitaries associated with noisy images not displaying a digit 0 or 1, in a third-class, this class will have measure 1, since the set of all unitaries that evolve the initial $\ket{\vec{0}}$ to some final pure product state $\ket{\phi}$ has Haar measure 0 in $\mathbb{SU}(N)$ \cite{lockhart_low-rank_2002}. For example when $n=1$, this can be seen by recognizing that the encoded states $\{\ket{\phi}\}$ correspond to only a fraction of the circle going through $\ket{0}$ and $\ket{1}$ on the Bloch Sphere. This labeling renders Theorem~\ref{thm: pc_risk_so} useless for any $h$ trained in this way because $\eta=0$. On the other hand, if we train a binary $h$ to classify half of $\mathbb{SU}(N)$, including unitaries corresponding to 0-digit images, to $l=0$, and the other half, including unitaries corresponding to 1-digit images, to $l=1$, then $\eta=1/2$. Using Eq.~\eqref{eqn: pc_risk_so} then gives $\mathcal{O}(1/\sqrt{d^n})$ robustness against prediction-change adversarial attacks, again suggesting extreme vulnerabilities in high dimensions.

However, the interpretation of this result is not of practical interest, for the following reasons. We emphasize that in applying Theorem~\ref{thm: pc_risk_so} or Eq.~\eqref{eqn:vul_epsilon}, the notion of adversarial risks by Definition~\ref{def:pc_risk_1} represents the probability of perturbing a Haar-randomly selected unitary by some $\epsilon$ to its adversarial example. It does not represent, for instance, the probability of perturbing a particular unitary associated with a real image to its adversarial example, nor does it represent the risk of attacking a unitary drawn from any other distribution over some subset. Therefore, if the task is to train and generalize a quantum classifier on a subset of quantum states with some distribution, this theorem cannot claim vulnerabilities that are exponential in the number of qudits. It is also noted that, as far as how Eq.~\eqref{eqn:vul_epsilon} and Theorem~\ref{thm: pc_risk_so} are formulated, the perturbed states cannot be mixed states since these are mapped from $\ket{\vec{0}}\bra{\vec{0}}$ by a completely positive and tracing preserving (CPTP) maps rather than by unitaries. In Section~\ref{sec:incorporation_generative}, we shall see that this is an example of an \textit{in-distribution} attack, which applies to scenarios where both the original and perturbed states are pure.

\section{Classifications on Generator Output Distributions} \label{sec:incorporation_generative}

\subsection{Concentration in Generated Distributions}
In practice, one is interested in the performance of a classifier on a distribution over some subset of meaningful samples, such as the subset of images displaying digits including the MNIST data set. It is this distribution on which the adversarial risk should be computed in order to infer the extent of the vulnerability. To ensure that the probability measure on the classifier-input space covers meaningful samples, we resort to approximating the distribution over meaningful samples using the image of a smooth generator function on a concentrated latent space, trained on samples of interest \cite{Fawzi2018}. Following convention, we refer to the latter as a real-data manifold. Such a generator can be a Normalizing Flow model \cite{Rezende,kingma_auto-encoding_2014,dinh_density_2017} or the generator of a Generative Adversarial Network (GAN) \cite{Goodfellow,bojanowski_optimizing_2018,arjovsky_wasserstein_2017}, both with a Gaussian latent space, trained on the same data set that the classifier will be trained on. A generative model serving this purpose is also referred to as a spanner \cite{Jalal2019}. In this way, a major fraction of the samples in the generator output $\mathcal{S}$ can be related to samples of interest, despite the fact that, the smoothness of the generator may introduce some samples off the real-data manifold, such as those undergoing gradual semantic change during interpolations. This generative set-up can be generalized to multiple generators on the same latent space. However, each generator maps to a disjoint part of the real-data manifold, overcoming the problem of covering the off real-data manifold when the latent space is globally connected \cite{khayatkhoei_disconnected_2018}. This generalization requires relaxing the demand that $\omega(0)=0$ in the Eq.~\eqref{eqn:modulus_continuity} below. As a result, no data off the real-data manifold is generated in $\mathcal{S}$. 

The reason that we require the latent space to be concentrated is so that we can study the concentration of samples in the generator-output space resulted from the concentration of the latent space. This connection is made by the assumption that the generator is smooth, in the sense that it admits a modulus of continuity (i.e., it is uniformly continuous), namely if there exists a monotone invertible function $\omega(\cdot)$ such that
\begin{equation} \label{eqn:modulus_continuity}
    \norm{g(z)-g(z')}\leq \omega(\norm{z-z'}_2),\quad \forall z, z'\in \mathcal{Z},
\end{equation}
 where $\norm{\cdot}$ is the metric equipped by the image of $g$. This is a weaker condition than the Lipschitz continuity which results when $\omega(\cdot)$ is a linear function. In this paper, we assume $\omega(\cdot)$ to yield a tight upper bound in Eq.~\eqref{eqn:modulus_continuity}, and we demand $\omega(\tau)$ to be small for small $\tau$ for a smooth generator. The idea is that any tendency to concentration of measure in the latent space is preserved by such a smooth mapping to its image, and the generated samples then follow a modified concentrated distribution. We can imagine that if some pairs of latent variables from different classes are within distance $b$ across the class boundary in the generator domain, their generator images must be accordingly within distance at most $\omega(b)$ across the boundary. This can also display a clustering. Although the tendency to cluster is preserved, the extent to which the points in the generator image gather is mediated by the modulus of continuity. A tight upper bound with $\omega(\cdot)$ that yields distances larger than the typical distances in the output space means that generated samples can be further apart, and vice versa. As far as adversarial robustness is concerned, a larger $\omega(\cdot)$ is then favorable since it implies that larger perturbations are needed to definitively perturb a larger number of generated samples across decision boundaries. 
 
 In generating these to-be-classified samples, the fact that a large probability density resides near the decision boundary is not at odds with a trained classifier that predicts training samples with high confidence. The training samples comprise only a subset of the support of the generator-output distribution. High confidence training samples result from the classifier drawing the decision boundaries away from them. When such a decision boundary encloses a sufficiently large measure, it then inevitably encounters large probability densities -- as dictated by the concentration of measure phenomenon on these distributions -- that do not contribute to training. As a result, when generalizing to test samples that are similar to the training samples, some test samples may locate near the boundary and be the vulnerable targets to adversarial attacks.

\subsection{Robustness of QML Models}\label{sec:qml_generator_quantum}

We consider the quantum adversarial attack set-up where the input to the classifier is already quantized and transmitted through a quantum communication network.

Let our latent space $\mathcal{Z}$ be, for example in this paper, the $\mathbb{R}^m$ with the Euclidean metric $\ell^2$ and the canonical $m$-dimensional Gaussian measure $\mathcal{N}_m\equiv\mathcal{N}(0,I_m)$ so it is a concentrated space. Let $z \leftarrow \mathcal{N}_m$ in $\mathcal{Z}$. Suppose that a smooth generator $g:\mathcal{Z}\rightarrow \mathcal{S}\subseteq\mathcal{X}$ is trained to generate a distribution $\xi$ of concern, such as some distribution of natural images, on a subset $\mathcal{S}$ of $\mathcal{X}$. For a sample $g(z)\in\mathcal{S}$, we then have $\xi(g(z))=\mathcal{N}_m(z)$.

 Incorporated in the generator $g=g_2\circ g_1$, $g_1$ maps the latent space to a subset of $n$-pixel natural images, $g_2$ then encodes the natural image into a density matrix defined in Eq.~\eqref{eqn:qudit}. That is, $g(z)=\ket{\phi(z)}\bra{\phi(z)}=\rho(z)\in\mathcal{S}\subseteq \mathcal{X}$, where $\mathcal{S}$ -- the image of $g$ -- is a subset of all density matrices $\mathcal{X}$. The metric on density matrices is the trace norm $L^1$ unless otherwise specified. The probability measure $\xi$, which is a distribution mapped by $g$ from the $m$-dimensional Gaussian measure $\mathcal{N}_m$ on $\mathcal{Z}$, is only supported on $\mathcal{S}$ over density matrices capturing the natural image distribution. Any quantum classifier $h$ then classifies the $d^n\times d^n$ density matrices in $(\mathcal{X},L^1,\xi)$. Let us denote the intermediate stage -- the set of images with $n$ pixels (normalized vectors with length n) -- as $\mathcal{I}$, then the corresponding measure on $\mathcal{I}$ can be denoted as $\xi\circ g_2$. The metric on $\mathcal{I}$ is, for instance, the $\ell^1$ norm. Diagrammatically, these mappings are
\begin{equation*}\label{eqn:generator}
    \mathcal{Z}\underbrace{\overset{g_1 }{\longrightarrow}\mathcal{I}\overset{g_2 }{\longrightarrow}}_{g}\mathcal{S}\subseteq\mathcal{X}\underset{h}{\longrightarrow}\mathcal{L}.
\end{equation*}

It is noted that smoothness is a desirable property of generative models. It is hinted at gradual transitions in the features in the generated samples, which imply that the generator has learned relevant factors of variation \cite{radford_unsupervised_2016}. We are then justified in assuming that the real-data manifold on $\mathcal{I}$ can be covered by a smooth generator $g_1$ (see e.g., \cite{kingma_auto-encoding_2014,dinh_density_2017,Goodfellow,bojanowski_optimizing_2018,arjovsky_wasserstein_2017}). In what follows, we show that the overall generator $g$, mapping from $\mathcal{Z}$ to the real-data manifold in the set of density matrices $\mathcal{X}$, is also smooth.

\begin{citedprop} \label{prop:g}
Assuming that $g_1:\mathcal{Z}\rightarrow\mathcal{I}$ is smooth with a modulus of continuity $\omega_1(\cdot)$ and the qudit encoding scheme, Eq.~\eqref{eqn:qudit}, is applied, then the generator $g=g_2\circ g_1:\mathcal{Z}\rightarrow\mathcal{S}\subseteq\mathcal{X}$ is also smooth and admits a modulus of continuity $\omega(\cdot)$ that is lower bounded as
\begin{equation*}
    \omega(\tau)\geq\sqrt{1-\cos^{2n(d-1)}\left(\frac{\pi}{2n}\omega_1(\tau)\right)},\quad \forall\tau\geq0.
\end{equation*}
\end{citedprop}
\noindent The proof can be found in Appendix~\ref{append:proof_omega_smooth}. In terms of the scaling with respect to $n$ and $d$, when $\omega_1(\cdot)$ scales as $\Omega(1)$, for instance, when $g_1$ is Lipschitz continuous (e.g., the generator in \cite{behrmann_invertible_2019,zhang_self-attention_2019}), \bww{Proposition~\ref{prop:g} implies that} the modulus of continuity of the overall generator $g$, \bww{i.e.,} $\omega(\cdot)$, scales as $\Omega(\sqrt{d/n})$. It is desirable to enforce Lipschitz continuity on some generators, for example when imposing spectral normalization \cite{miyato_spectral_2018} on the generator of a GAN to improve training \cite{zhang_self-attention_2019}.
 
\par A distinction can be made concerning whether the adversarial example $\sigma$ must be also in the subset $\mathcal{S}$. If so, the adversarial attack is called in-distribution, since the attacker only looks for an adversarial example within the data manifold $\mathcal{S}$. Otherwise, we call it an unconstrained adversarial attack since the perturbation is arbitrary in $\mathcal{X}$, i.e., it is not confined to the data manifold. We state the precise definitions, based on prediction-change adversarial risks in Definition~\ref{def:pc_risk_1}, as follows.

\begin{definition} \label{def:in_dist_risk}
An in-distribution adversarial attack, or a data-manifold attack, attempts to find the perturbation
\begin{equation*}
\begin{split}
\varepsilon_{in}(\rho)&=\min_{r\in\mathcal{Z}}\{\norm{g(z+r)-\rho}_1| h(g(z+r))\neq h(\rho)\}\\
&=\min_{\sigma\in\mathcal{S}}\{\norm{\sigma-\rho}_1|h(\sigma)\neq h(\rho)\},
\end{split}
\end{equation*}
which is within the data manifold $(\mathcal{S}, L^1, \xi)$. It induces an in-distribution adversarial risk,
\begin{equation*}
   R^{PC}_{\epsilon_{in}}(h,\xi)=\Pr_{\rho\leftarrow\xi}\left[\varepsilon_{in}(\rho)\leq\epsilon_{in}\right].
\end{equation*}
\end{definition}

\begin{definition} \label{def:unc_risk}
An unconstrained adversarial attack attempts to find
\begin{equation*}
\varepsilon_{unc}(\rho)=\min_{\sigma\in\mathcal{X}}\{\norm{\sigma-\rho}_1|h(\sigma)\neq h(\rho)\},
\end{equation*}
which is in $(\mathcal{X}, L^1)$ not restricted to the data manifold $\mathcal{S}$. It induces an unconstrained adversarial risk,
\begin{equation*}
    R^{PC}_{\epsilon_{unc}}(h,\xi)\equiv R^{PC}_{\epsilon}(h,\xi)=\Pr_{\rho\leftarrow\xi}\left[\varepsilon_{unc}(\rho)\leq\epsilon\right].
\end{equation*}
\end{definition}

 It is noted that when the generator is surjective on $\mathcal{X}$, i.e., $\mathcal{S}=\mathcal{X}$, there is no distinction between the two types of attacks.The set-ups in Theorem~\ref{thm: pc_risk_so} and Eq.~\eqref{eqn:vul_epsilon} consider classifying on the subset of all pure-state density matrices in $\mathcal{X}$ on which a Haar-random distribution $\nu$ is supported. Since this requires both the original and perturbed states be pure, the adversarial risks are considered in-distribution, although we shall see in Section~\ref{sec:unconstrained_robustness} that the same upper bound applies to the unconstrained robustness for a general quantum classifier.

\subsubsection{In-distribution Adversarial Robustness} 
 The following theorem, depending on the distribution to be classified as well as the specific classical-data generator $g_1$ in terms of $\omega_1(\cdot)$, then applies.

\begin{theorem}\label{thm: in_dist_risk}
Let $h: \mathcal{X}\rightarrow \mathcal{L}$ be any quantum classifier on the set of density matrices. 
Considering in-distribution adversarial attacks on the image of $g$,
if $\xi(h^l))\leq 1/2, \forall l$, i.e., the classes are not too unbalanced, then for the prediction-change risk $R^{PC}_{\epsilon_{in}}(h, \xi)\leq1-\gamma$, the distance between two density matrices $\epsilon_{in}$ must satisfy

\begin{equation} \label{eqn:gen_eps}
\epsilon_{in} \leq \omega\left(\sqrt{\ln{\left(\frac{\pi}{2\gamma^2}\right)}} \right),
\end{equation}
\bww{where $\omega(\cdot)$ is the modulus of continuity} in Proposition~\ref{prop:g}. 
\end{theorem}

\noindent The proof can be found in Appendix~\ref{append:proof_major_thm}. This result is independent of the quantum data encoding scheme. It can be generalized to quantum classifiers with arbitrary decision boundaries, but in this case, the necessary upper bound on the in-distribution robustness will not have a closed-form (see Appendix~\ref{append:proof_major_thm}). This upper bound is saturated when Eq.~\eqref{eqn:modulus_continuity} is tight and the quantum classifier induces linearly separable regions in the latent space, namely when $h\circ g$ is a linear function on $\mathcal{Z}$, giving rise to the maximally robust quantum classifier. The non-saturation of this upper bound when class regions are not linearly separable in $\mathcal{Z}$ can be seen in the example of the standard Gaussian in Section~\ref{sec:concentration_of_measure} above. Suppose one looks at $\Sigma'=(-\infty,-2\delta)\cup(0,2\delta)$ for some $\delta>0$, which has the same probability measure $1/2$ as $\Sigma=(-\infty,0)$ but is not linearly separable in $\mathbb{R}$. The measure outside the $\delta$-expansion of $\Sigma'$, i.e., $\mathbb{R}\setminus\Sigma'_\delta=(3\delta,+\infty)$, is smaller than that outside of the $\delta$-expansion of $\Sigma$, namely $\mathbb{R}\setminus\Sigma_\delta=(\delta,+\infty)$, implying more concentration outside and near $\Sigma'$ than $\Sigma$. 

The non-saturation of this upper bound for non-linearly separable classification regions in $\mathcal{Z}$ also implies that it is prone to misclassification with an increasing number of equiprobable classes. The proof for cases with at least 5 equiprobable classes can be found in Appendix~\ref{append:proof_major_thm}. Informally, more equiprobable classes lead to more boundaries, enclosing classes with sufficiently large total measure, that border distinct classes. Then within a fixed distance beyond more of those boundaries, there are more samples subject to some prediction change. 

We note that this upper bound is usually not saturated in practice, since a quantum classifier is usually linear, such as a parametrized quantum circuit and a unitary tensor network, while the generator $g$ is usually non-linear, given that $g_1$ is usually non-linear and $g_2$, the quantum feature map, is non-linear. Classically, some highly-nonlinear state-of-the-art neural networks have robustness one or two orders of magnitude smaller in the $\ell^2$ norm on some data sets than the corresponding upper bound derived with similar arguments \cite{Fawzi2018}. It would be interesting to examine the amount of deviation from the upper bound for QML models in future works. 

Theorem~\ref{thm: in_dist_risk} \bww{implies} that
when the \bww{quantum states to be classified} encode classical data generated with a modulus of continuity \bww{scaling} as $\Omega(1)$, the in-distribution robustness of any quantum classifier decreases polynomially in the number of qudits \bww{$n$} and increases polynomially in the qudit dimension \bww{$d$}. \bww{To see this, we first note that according to Proposition~\ref{prop:g}, when $\omega_1(\cdot)=\Omega(1)$, which applies to} generators such as those enforcing Lipschitz continuity, $\omega(\cdot)$ is lower bounded by a function that scales as $\Omega(\sqrt{d/n})$. \bww{This means that the upper bound on the perturbation size $\epsilon_{in}$ between any two in-distribution states, i.e., the right hand side of Eq.~\eqref{eqn:gen_eps}, is then asymptotically bounded from below by $\sqrt{d/n}$.}

As such, the vulnerability increases slightly with a larger number of qudits $n$ and by contrast, decreases slightly with qudits of higher dimension $d\geq2$. When the encoded classical data manifold comes from generators for which Lipschitz continuity is not enforced, it requires numerical approximations of the modulus of continuity $\omega_1(\cdot)$ to determine its scaling in the output space, before obtaining the robustness scaling. Compared to Theorem~\ref{thm: pc_risk_so} where samples are Haar-random pure states, the states to be classified here, which characterise the adversarial risk, are similar to those considered in practical tasks. Specifically, they are a subset of encoded states with a distribution smoothly generated from a Gaussian latent space. Theorem~\ref{thm: in_dist_risk} demonstrated that, contrary to previous claims \cite{Liu}, there is no guarantee that quantum classifiers are exponentially more vulnerable to in-distribution attacks in higher-dimensional Hilbert space. \bww{We shall now show} that the theorem applies to unconstrained attacks as well.

\subsubsection{Unconstrained Adversarial Robustness}\label{sec:unconstrained_robustness}
Unconstrained adversarial attacks are arbitrary perturbations in $\mathcal{X}$ to a sample $\rho$. In a broader context in which the instability of the quantum classifier is concerned, this may derive from density matrices subject to decoherence in a classification task. It is clear that $\varepsilon_{unc}(\rho)\leq\varepsilon_{in}(\rho), \forall \rho\in\mathcal{X}$ and thus, it holds by changing the in-distribution perturbations in Theorem \ref{thm: in_dist_risk} to unconstrained ones, and the same bound in Eq.~\eqref{eqn:gen_eps} applies. 

We argue that there does not exist a tighter upper bound that holds for general quantum classifiers for unconstrained robustness. Consider a particular family of quantum classifiers that project any state onto the data manifold, namely to map any state to its closest in-distribution state, before classifying. These classifiers can be shown to satisfy $1/2\varepsilon_{in}(\rho)\leq\varepsilon_{unc}(\rho)\leq\varepsilon_{in}(\rho),\forall\rho\in\mathcal{X}$ \footnote{It is proven in Theorem 2 in \cite{Fawzi2018}.}. Even in the worst case where $\varepsilon_{unc}(\rho)=1/2\varepsilon_{in}(\rho),\forall \rho\in\mathcal{X}$, their unconstrained robustness is as large as half of the in-distribution one. We stress that, although robust, such a quantum classifier is inefficient in our setting since there is no apparent tractable way to obtain the closest pure product state to an arbitrary state.

\begin{table*}[htbp]
\centering
\begin{tabular}{ c|c c } 
 \hline
 $\leq1-\gamma$ & $\norm{\rho-\sigma}_1\leq$ & $\norm{u-v}_1\leq$\\ 
 \hline
 \multirow{3}{*}{$R^{PC}_{\epsilon}(h,\nu)$} & \multirow{3}{*}{$4d^{-n}\lambda_1=\Omega(d^{-n})$} & 
 \multirow{3}{*}{$\frac{2n}{\pi}\cos^{-1}\left[\left(1-2d^{-n}\lambda_1\right)^{\frac{1}{(d-1)n}}\right]=\Omega(d^{-\frac{n}{2}}\sqrt{n})$} \\
 & & \\
 & & \\
  \multirow{3}{*}{\shortstack{$R^{PC}_\epsilon(h,\xi)$}}
  & \multirow{3}{*}{\quad$\omega(\lambda_2)\geq\sqrt{1-\cos^{2n(d-1)}\left(\frac{\pi}{2n}\omega_1\left(\lambda_2\right)\right)}=\Omega\left(\sqrt{\frac{d}{n}}\right)\quad$} & \multirow{3}{*}{$\omega_1\left(\lambda_2\right)=\Omega(1)$} \\ 
  & & \\
  & & \\
 \hline
\end{tabular}
\caption{\bww{Summary of the adversarial robustness, \hl{namely the size of perturbations necessary for the adversarial risk to be upper bounded by some constant,} of any quantum classifier obtained within the prediction-change adversarial attack setting. In this setting, the prediction-change adversarial risk over the Haar-random distribution $\nu$ ($R^{PC}_{\epsilon}(h,\nu)$) and over a smoothly generated distribution $\xi$ ($R^{PC}_\epsilon(h,\xi)$) are both upper bounded by $(1-\gamma)$ (column 0). $d$ denotes the qudit dimension in Eq.~\eqref{eqn:qudit} and $n$ denotes the number of encoded qudits or the length of the encoding vectors (number of pixels in the image classification example). Parameters $\lambda_1$ and $\lambda_2$ are defined as $\lambda_1=[\ln(2\sqrt{2}/\eta)]^{1/2}+[\ln(2\sqrt{2}/\gamma)]^{1/2}$ and  $\lambda_2=\sqrt{\ln{(\pi/(2\gamma^2))}}$. Row 1 summarizes the adversarial robustness when a pure state $\rho$ sampled from the Haar-random distribution $\nu$ is perturbed to a state $\sigma$. The robustness is shown both in the trace norm (column 1), as well as in its translation to the robustness measured in the $\ell^1$ norm of the set of encoding vectors (from Corollary \ref{cor:to_thm_pc_risk} of Theorem~\ref{thm: pc_risk_so}) (column 2). Both upper bounds decrease exponentially in $n$. Row 2 summarizes the adversarial robustness when a pure state $\rho$ sampled from a smoothly generated distribution $\xi$ from a Gaussian latent space is perturbed to a state $\sigma$ (column 1), and the robustness when the intermediately generated vector $u$ is perturbed to $v$ \hl{(column 2)} \hl{(from Proposition~\ref{prop:g} and Theorem~\ref{thm: in_dist_risk})}. Note that when the robustness in adversarially perturbing a vector scales as $\Omega(1)$, \hl{e.g., when the intermediate vectors are generated Lipschitz continuously,} that in perturbing an encoded pure state scales as $\Omega(\sqrt{d/n})$.}
}
\label{table:1}
\end{table*}

Inspired by this strategy, we propose that one can construct a family of efficient quantum classifiers $\tilde{h}$ on $n$-qubit density matrices $\mathcal{X}$ with unconstrained robustness $\varepsilon_{unc}(\rho)$ lower bounded for any $\rho\in\mathcal{X}$. To be specific, we construct $\tilde{h}$ from any $h$ with the following procedure.

Let the original sample $\rho\in\mathcal{S}$ be a pure product-state density matrix with $n$ qudits as in Eq.~\eqref{eqn: product_states}. A perturbation $\epsilon_{unc}\equiv\epsilon$ leads to a sample $\sigma\in\mathcal{X}$. First, we perform single qubit tomography on every qubit of $\sigma$ to reconstruct a product-state density matrix from these single qubits. We denote this mapping as $P: \mathcal{X}\rightarrow\mathcal{X},\ \sigma \mapsto \bigotimes_{i=1}^n\tr_{\{j\neq i\}}(\sigma)$. Subsequently, we numerically fit the pixel values $\{s_i\}$ to $P(\sigma)$ to find its closest density matrix $\tilde{\sigma}$ within our data manifold $\mathcal{S}$. We use a symbol $\tilde{\sigma}$ to represent the density matrix attained from this procedure. $\tilde{\sigma}$ is then replacing $\sigma$ when fed to the quantum classifier $h$. We have the following theorem,
\begin{theorem} \label{thm:sqt}
    For every $n$-qubit $\rho\in \mathcal{S}\subseteq\mathcal{X}$, let $\tilde{\rho}$ be the density matrix within the data manifold attained from the above procedure.
    For any quantum classifier $h$, let $\tilde{h}:\mathcal{X}\rightarrow\mathcal{L}$ be such that $\tilde{h}(\rho)=h(\tilde{\rho})$, then 
 \begin{equation*}
    2-2\left(1-\frac{\varepsilon_{in}(\rho)^2}{16}\right)^{\frac{1}{n_e}}\leq \varepsilon_{unc}(\rho)\leq \varepsilon_{in}(\rho),
\end{equation*}
where $n_e=n$ for even $n$ and $n_e=n+1$ for odd $n$.
\end{theorem}
\noindent The proof can be found in Appendix~\ref{append:proof_thm_sqt}. We note that the procedure can be applied to any product state encoding scheme. This procedure yields an explicit lower bound to the unconstrained adversarial perturbation when it is possible to estimate the in-distribution adversarial perturbation by, for example, sampling in the latent space \cite{zhao_generating_2018} or gradient descent search in the latent space \cite{Jalal2019} before mapping to the density matrices. This $\tilde{h}$ constructed from $h$ amounts to a feasible tomographic preprocessing of input states. It guarantees that the unconstrained robustness of each sample $\rho$ is bounded from below and may be used as a defense strategy against unconstrained adversarial attacks in practice. However, we note that when $n$ is large, this lower bound can be several orders of magnitude smaller than the upper bound.

\section{Discussion} \label{sec:comparison}
\bww{A summary of the upper bounds on the \hl{prediction-change adversarial robustness} over \hl{pure states sampled from} the Haar-random distribution $\nu$ and a smoothly generated distribution $\xi$, is presented in Table~\ref{table:1}.}

\hl{In this work, we first showed the prediction-change adversarial robustness over Haar-randomly distributed pure states, and compared this with the previously demonstrated error-region robustness of \cite{Liu} over the same distribution. Both types of adversarial robustness show similar extreme vulnerabilities exponential in the number of qudits.}
However, \bww{in this work we have \hl{argued} that \hl{these vulnerabilities} for Haar-random pure states} \hl{are} not of practical interest. \correct{This is because, in practice, the adversarial risk of a quantum classifier should be computed on a distribution over some subset of meaningful states, such as a subset of qubit encoding states featurizing some images, \bw{in order} to infer the extent of the vulnerability}. \bw{In general,} \correct{practical quantum classification tasks classify a subset of encoded states with some commonly used qubit encoding scheme. \bw{For such} tasks, we \bww{have shown that we can} use the concentration of measure phenomenon to derive the robustness of any quantum classifiers} in situations where the \hl{distribution of states to be classified can be smoothly generated from a Gaussian latent space}, \bw{as quantified in} Eq.~\eqref{eqn:modulus_continuity}. \bww{In this situation, we have shown that one finds only a mildly polynomially decreasing robustness in the number of such encoded qubits, specifically with scaling} as $\mathcal{O}(\sqrt{1/n})$ in the trace norm.

\bww{As noted for Theorem~\ref{thm: in_dist_risk}, it is the upper bound on the perturbation size necessary for the adversarial risk to be bounded from above that scales as $\Omega(\sqrt{1/n})$. This upper bound is usually not tight and the actual adversarial robustness could therefore be smaller. We have also proposed} a feasible modification of any quantum classifier \correct{with product-state inputs} -- \bww{namely,} by performing single qubit tomography before numerically fitting the closest encoded qubit state -- to obtain a lower bound on the unconstrained robustness and to defend against unconstrained adversarial attacks.

\correct{Most importantly, our analysis provides QML protocols some relief from adversarial attacks in real-world tasks. For example, when classifying on some qubit states \hl{encoding} MNIST images, the robustness decreases only as $\mathcal{O}(\sqrt{1/n})$, in contrast to the extreme vulnerability of quantum classifiers in classifying Haar-random pure states (Theorem~\ref{thm: pc_risk_so} and \cite{Liu}). In future, it will be interesting to experimentally compare the adversarial robustness of particular QML models \bw{for real-world data} on a distribution of states smoothly mapped from a Gaussian latent space with the bounds \bw{that} we have derived \bw{here}.}

We note that the polynomially decreasing robustness in $n$ is \bw{derived} from the qudit encoding scheme. The concentration of measure due to the Gaussian isoperimetric inequality for the latent space only contributes to the argument of Eq.~\eqref{eqn:gen_eps}. It will be interesting to investigate \bw{whether} a different encoding scheme can give better scaling in the robustness, and \bw{also to determine whether} quantum data that \bw{derives} naturally from a distribution other than the Haar-random distribution is robust to attacks. In Appendix~\ref{append:adv_TTN}, we propose a method to perform white-box adversarial attacks on classically intractable input states \bw{with} QML models. It will be interesting to further explore white-box attacks, assuming \bw{that} the adversary is capable of devising these. In practice, with current NISQ-era hardware, it will \bw{also be useful} to examine how robust QML models are against adversarial attacks under noise and decoherence.

\section*{Acknowledgement} \label{sec:acknowledgements}
H. L. was supported by the National Aeronautics and Space Administration under Grant/Contract/Agreement No.80NSSC19K1123 issued through the Aeronautics Research Mission Directorate. I.C. was supported by the US Department of Energy, Office of Science, Office of Advanced Scientific Computing Research, Quantum Algorithm Teams Program, under contract number DE-AC02-05CH11231.   W. H. was supported by a grant from Siemens Corporation.

%

\section*{Appendices}
\begin{appendix}
\section{Confidence Difference and Distance between States} \label{append:metric}
We show that the predictive confidence difference in any QML protocol is upper bounded by the distance between the input density matrices up to some constant factor, \bww{where this distance is measured in the trace norm $L^1$, the Hilbert-Schmidt norm $L^2$, the Bures distance, or} the Hellinger distance.

Considering density matrices $\rho$ and $\sigma$, the trace norm between them is defined to be $\norm{\rho-\sigma}_1=\tr(\abs{\rho-\sigma})$. Consider a set of POVMs $\{\Pi_l\}$ and a quantum channel $\mathcal{E}$ such that $\mathcal{E}(\rho)=\sum_i M_i \rho M_i^\dagger$ and $\sum_i M_i^\dagger M_i=I$. We have,
\begin{equation*}
    \begin{split}
        \tr(\mathcal{E}(\rho)\Pi_l)-\tr(\mathcal{E}(\sigma)\Pi_l)&=\tr\left(\sum_i M_i (\rho-\sigma) M_i^\dagger \Pi_l\right)\\
        &=\tr\left((\rho-\sigma) \sum_i M_i^\dagger \Pi_l M_i\right)\\
        &\equiv\tr((\rho-\sigma) \mathcal{E}^*(\Pi_l)).
    \end{split}
\end{equation*}
We note that $\mathcal{E}^*$ is the dual map of $\mathcal{E}$ and $\{\mathcal{E}^*(\Pi_l)\}$ is still a set of POVMs, since $\mathcal{E}^*(\Pi_l)$ is hermitian, non-negative because $\tr (\rho\mathcal{E}^*(\Pi_l) )=\tr (\mathcal{E}(\rho)\Pi_l) \geq 0$, and complete because $\sum_{i,s}M_i^\dagger \Pi_l M_i = \sum_i M_i^\dagger M_i = I$. 

For each particular measurement, we can expand in its eigenbasis $\mathcal{E}^*(\Pi_l)=\sum_k b_k \ket{\phi_k}\bra{\phi_k}\equiv\sum_k b_k P_k$. Let $\{\ket{\psi_i}\}$ and $\{\lambda_i\}$ be the eigenbasis and eigenvalues of $(\rho-\sigma)$, so
$\norm{\rho-\sigma}_1 = \sum_i |\lambda_i|\in[0,2]$.
We then expand $\mathcal{E}^*(\Pi_l)=\sum_{i,j,k} b_k   a_{ik}\ket{\psi_i}a^{*}_{jk}\bra{\psi_j}$ such that $\sum_{i}\abs{a_{ik}}^{2}=1,{\forall} {k}$ and $\sum_k b_k=\tr(\mathcal{E}^*(\Pi_l))\geq 0$ due to the non-negativity. We have
\begin{equation}\label{eqn:1}
\begin{split}
    \tr((\rho-\sigma)\mathcal{E}^*(\Pi_l&)) = \tr\left((\rho-\sigma)\sum_{i,j,k}b_k a_{ik}\ket{\psi_i}a^*_{jk}\bra{\psi_j}\right)\\
    &= \sum_k b_k \tr\left(\sum_{i,j} a_{ik} a_{jk}^* \bra{\psi_j} (\rho-\sigma) \ket{\psi_i} \right)\\
    &= \sum_{i,k} b_k |a_{ik}|^2 \lambda_i\leq \sum_k b_k \norm{\rho-\sigma}_1\\
    &= \tr({\mathcal{E}^*(\Pi_l)})\norm{\rho-\sigma}_1.
\end{split}
\end{equation}
Therefore,
\begin{equation*}
    \abs{\tr(\mathcal{E}(\rho)\Pi_l)-\tr(\mathcal{E}(\sigma)\Pi_l)} \leq \tr({\mathcal{E}^*(\Pi_l)}) \norm{\rho-\sigma}_1.
\end{equation*}
When $\tr({\mathcal{E}^*(\Pi_l)})$ is \bww{not too large} the above inequality suggests that the confidence change \bww{will} be small when the trace norm between the two density matrices is small. However, $\tr({\mathcal{E}^*(\Pi_l)})$ may be very large in high dimensions and in that case, the upper bound becomes very weak. 

We resort \bww{instead} to the physical interpretation of trace distance being a generalization of the classical total variation distance. The trace distance between two quantum states is an achievable upper bound on the total variation distance between probability distributions arising from measurements performed on those states \cite{nielsen_quantum_2010}:
\begin{equation*}
    \frac{1}{2}\norm{\rho-\sigma}_1=\frac{1}{2}\max_{\{\Pi_l\}}\sum_l\abs{\tr[(\rho-\sigma)\Pi_l]},
\end{equation*}
where the maximization is over all POVMs $\{\Pi_l\}$ \correct{and the factor of $2$ is to restrict the maximal trace distance to be $1$}. Using the contractive property of the trace norm under any CPTP map, we conclude that the trace norm \bww{constitutes} an upper bound to the sum of confidence changes of all POVMs:
\begin{equation}\label{eqn:trace_norm_upper_bound}
   \sum_l\abs{\tr(\mathcal{E}(\rho-\sigma)\Pi_l)} \leq \norm{\mathcal{E}(\rho)-\mathcal{E}(\sigma)}_1 \leq 
   \norm{\rho-\sigma}_1.
\end{equation}

\bww{Considering} the Hilbert-Schmidt norm defined as $\norm{\rho-\sigma}_2^2=\tr[(\rho-\sigma)^2]$, if we regard $\norm{\rho-\sigma}_2$ as the inner product of the two vectors $(1, 1, \cdots, 1)$ and $(\abs{\lambda_0}, \abs{\lambda_1}, \cdots, \abs{\lambda_{N-1}})$, then from the Cauchy-Schwarz inequality we find $\norm{\rho-\sigma}_1\leq\sqrt{N}\norm{\rho-\sigma}_2$. But this bound is very weak in high dimensional Hilbert space. A better upper bound is given in \cite{coles_strong_2019} that $\norm{\rho-\sigma}_1\leq 2\sqrt{R}\norm{\rho-\sigma}_2$, where $R=\text{rank}(\rho)\text{rank}(\sigma)/[\text{rank}(\rho)+\text{rank}(\sigma)]$. This implies that, even
when one state is full rank, if the other state is low rank,
then the Hilbert-Schmidt norm is of the same order of
magnitude as the trace norm. This is the case when we consider any perturbation to an encoded pure state density matrix $\rho$ whose rank is $1$. Combined with Eq.~\eqref{eqn:trace_norm_upper_bound}, we arrive at a similar upper bound,
\begin{equation*}
    \sum_l\abs{\tr(\mathcal{E}(\rho)\Pi_l)-\tr(\mathcal{E}(\sigma)\Pi_l)} \leq 2\sqrt{R}\norm{\rho-\sigma}_2.
\end{equation*}

$\ $ \bww{Considering} the Bures distance defined as $\norm{\rho-\sigma}_B^2=2(1-\sqrt{F(\rho, \sigma)})$, it is an extension to mixed states of the Fubini-Study distance for pure states \cite{Spehner2017}. We have
\begin{equation*}
\begin{split}
    \norm{\rho&-\sigma}_1\leq 2\sqrt{1-\left(1-\frac{1}{2}\norm{\rho-\sigma}_B^2\right)^2}\\
    &= 2\sqrt{\norm{\rho-\sigma}_B^2-\frac{1}{4}\norm{\rho-\sigma}_B^4}\leq2\norm{\rho-\sigma}_B,
\end{split}
\end{equation*}
where the first inequality is proven in \cite{s_holevo_quasiequivalence_1972, Spehner2017} and saturated for pure states. Therefore, together with Eq.~\eqref{eqn:trace_norm_upper_bound}, we conclude that
\begin{equation}\label{eqn:bures_distance}
   \sum_l\left|\tr(\mathcal{E}(\rho)\Pi_l)-\tr(\mathcal{E}(\sigma)\Pi_l)\right| \leq 2\norm{\rho-\sigma}_B.
\end{equation}

\bww{Finally, considering} the Hellinger distance defined as $\norm{\rho-\sigma}_H^2=2-2\tr(\sqrt{\rho}\sqrt{\sigma})$, it is shown that $\norm{\rho-\sigma}_B\leq\norm{\rho-\sigma}_H$ \cite{Spehner2017} and thus, the same upper bound applies by changing $\norm{\rho-\sigma}_B$ to $\norm{\rho-\sigma}_H$ in Eq.~\eqref{eqn:bures_distance}.

In QML, if $\rho$ and $\sigma$ are close in these norms and \bww{are separated by} any class boundary, say between class $l=s$ and class $l=t$, then $\tr(\mathcal{E}(\rho)\Pi_s)>\tr(\mathcal{E}(\sigma)\Pi_s)$ while $\tr(\mathcal{E}(\rho)\Pi_t)<\tr(\mathcal{E}(\sigma)\Pi_t)$. \bww{This} suggests that no small perturbation to density matrices in these norms can significantly change the measurement outcome and thus, alter the prediction, unless the original sample is near the boundary. In other words, viewing $\tr(\mathcal{E}(\rho)\Pi_s)$ as the confidence of predicting $l=s$, it implies that no small perturbations can result in a high-confidence sample in one class perturbed to a low-confidence sample in the same class, or a high-confidence sample in a different class. 

\section{Adversarial Attacks Exploiting Quantum Classifier Reversibility}\label{append:adv_TTN}
We propose a method to perform adversarial attacks in our first set-up in Section~\ref{sec:qml_attack_setup} on quantized data. This method can be carried out on a quantum hardware when the computation is classically intractable. We assume a noiseless QML model for this analysis, so the quantum channel is unitary. Considering, for example, the unitary tree tensor network (TTN) in \cite{Huggins2019} among other types of parametrized unitary quantum circuits, the adversary can run it reversely starting from a density matrix with any designated wrong class label $l=t$ such that $\tr(\sigma'\Pi_t)=1$ while $\tr(\sigma'\Pi_{l\neq t})=0$. Any qubit that is traced out in the forward direction is initialized to an arbitrary state and passes through the network in the reverse direction. The output of the reversal circuit is a set of density matrices $\{U^\dagger\sigma' U\}\equiv\{\sigma\}$ such that $\tr(U \sigma U^\dagger\Pi_t)=1$ whereas $\tr(U \sigma U^\dagger\Pi_{l\neq t})=0$. Thus, this set of density matrices will be classified in the wrong class by the POVM $\Pi_t$ with high-confidence. 
  Suppose that the original samples are $\{\rho\}$ in the class $s\neq t$ and $\tr(U\rho U^\dagger\Pi_s)=1/2+\delta$ with some $\delta\in(0,1/2]$. The adversary then replaces an $\epsilon$-portion of the transmitted quantum states $\{\rho\}$ with the $\{\sigma\}$ to attack the receiver.

To achieve a prediction change, the adversary demands $\tr(U[(1-\epsilon)\rho+\epsilon\sigma]U^\dagger\Pi_s)<1/2$. This requires
\begin{equation}\label{eqn:substitution_attack}
    \epsilon>1-\frac{1}{1+2\delta},
\end{equation}
which means that the portion of $\{\rho\}$ being substituted with $\{\sigma\}$ increases with higher-confidence of $\{\rho\}$. We note that this effectively creates a perturbation of size
\begin{equation*}
    \begin{split}
        \norm{\rho-[(1&-\epsilon)\rho+\epsilon \sigma]}_1\geq\epsilon\sum_l\abs{\tr(U(\rho-\sigma)U^\dagger\Pi_l)}\\
        &=\epsilon\Big[\sum_{l\neq t}\tr(U\rho U^\dagger\Pi_l)+(1-\tr(U\rho U^\dagger\Pi_t))\Big]\\
        &=\epsilon[2-2\tr(U\rho U^\dagger\Pi_t)]\geq\epsilon(1+2\delta),
    \end{split}
\end{equation*}
where the first inequality follows from Eq.~\eqref{eqn:trace_norm_upper_bound}. As a result, a misclassification by the attack demands a perturbation of size $\norm{\rho-[(1-\epsilon)\rho+\epsilon \sigma]}_1\geq2\delta$ where we substituted in Eq.~\eqref{eqn:substitution_attack}.

\section{Proof of Eq.~(\ref{eqn:vul_epsilon})}\label{append:nana_correction}
We present a condensed proof based on the proof to Theorem 3.7 in \cite{Mahloujifar2019}. Let $\epsilon_1>\sqrt{1/(Nk_2)\ln{(k_1/\mu(\mathcal{M}))}}$ and $\epsilon_2> \sqrt{1/(Nk_2)\ln{(k_1/\gamma)}}$. Then the concentration function satisfies $\alpha(\epsilon_1)<\mu(\mathcal{M})$ and $\alpha(\epsilon_2)<\gamma$. As such, by directly applying Part 2 of Theorem 3.2 in \cite{Mahloujifar2019}, we conclude $R^{ER}_\epsilon(h,c,\nu)>1-\gamma$ for $\epsilon=\epsilon_1+\epsilon_2$. It can be shown that $\mathbb{SU}(N)$ is a $(\sqrt{2}, 1/4)$-normal L\'evy family and so $k_1=\sqrt{2}$ and $k_2=1/4$ \cite{Liu}. 
The contrapositive statement on $R^{ER}_\epsilon(h,c,\nu)\leq1-\gamma$ then gives the necessary condition Eq.~\eqref{eqn:vul_epsilon}.

\section{Proof of Theorem~\ref{thm: pc_risk_so}} \label{append:proof_thm_pc_risk_su}
\begin{proof}
We let $\epsilon_1>\sqrt{1/(Nk_2)\ln{(2k_1/\eta)}}$ and $\epsilon_2> \sqrt{1/(Nk_2)\ln{(2k_1/\gamma)}}$, then the concentration function satisfies $\alpha(\epsilon_1)<\eta/2$ and $\alpha(\epsilon_2)<\gamma/2$. Therefore, by applying Part 1 of Theorem A.2 in \cite{Mahloujifar2019}, we conclude that for $\epsilon=\epsilon_1+\epsilon_2$, $R^{PC}_\epsilon(h, \nu)>1-\gamma$. For completeness, we present our explained version of the proof below.

Let $\epsilon=\epsilon_1+\epsilon_2$. By assumption that $\nu(h^l)\leq1-\eta, \forall l\in\mathcal{L}$, it can be easily verified by contradiction that $\exists l_1\in\mathcal{L}$ s.t. $\nu(h^{l_1})\in(\eta/2,1/2]$. Let $h^{l_1,C}=\mathcal{X}\setminus h^{l_1}$. On one hand, we know that $\nu(h^{l_1})>\eta/2> \alpha(\epsilon_1)$ where the last inequality is given by our assumption. We prove by contradiction that
$\nu(h^{l_1}_{\epsilon_1})>1/2$. Suppose not, then we have for $\mathcal{S}=\mathcal{X}\setminus h^{l_1}_{\epsilon_1}$, $\nu(\mathcal{S})=1-\nu(h^{l_1}_{\epsilon_1})\geq1/2$. Then by the definition of the concentration function in Eq.~\eqref{eqn:concentration_func}, $\nu(\mathcal{S}_{\epsilon_1})\geq1-\alpha(\epsilon_1)$. Combining with what we obtained that $\nu(h^{l_1})>\alpha(\epsilon_1)$, we have $\nu(\mathcal{S}_{\epsilon_1})+\nu(h^{l_1})>1$. Thus, $\exists x\in \nu(\mathcal{S}_{\epsilon_1})\cup\nu(h^{l_1})$. This implies $\exists y\in\mathcal{S}|\mathrm{d}(y,x)\leq \epsilon_1$. But this same $y$ must also be in $h^{l_1}_{\epsilon_1}$ since the same $x$ is also in $h^{l_1}$. However, this raises a contradiction since $\mathcal{S}$ and $h^{l_1}_{\epsilon_1}$ are disjoint by definition, i.e., $\nexists y|y\in\mathcal{S}, y\in h^{l_1}_{\epsilon_1}$. 
Now, $\nu(h^{l_1}_{\epsilon_1})>1/2$ means, by the definition of the concentration function in Eq.~\eqref{eqn:concentration_func}, as well as the assumption that $\gamma/2>\alpha(\epsilon_2)$, we have $\nu(h^{l_1}_{\epsilon})\geq1-\alpha(\epsilon_2)>1-\gamma/2$.

On the other hand, knowing that $\nu(h^{l_1,C})\geq1/2$, we have that $\nu(h^{l_1,C}_{\epsilon_2})>1-\gamma/2$ followed by simply replacing the $h^{l_1}_{\epsilon_1}$ in the previous sentence with $h^{l_1,C}$ since they both have measure at least $1/2$. We then also have $\nu(h^{l_1,C}_{\epsilon})>1-\gamma/2$. Hence, using the inequality $\mu(\cap_{i=1}^n A_i)\geq\sum_{i=1}^n\mu(A_i)-(n-1)$, one can conclude that $\nu(h^{l_1}_\epsilon\cap h^{l_1,C}_\epsilon)>1-\gamma$ and so, by the prediction-change risk's definition, $R^{PC}_\epsilon(h,\nu)\geq\nu(h^{l_1}_\epsilon\cap h^{l_1,C}_\epsilon)>1-\gamma$.

It can be shown that $\mathbb{SU}(N)$ is a $(\sqrt{2}, 1/4)$-normal L\'evy family and so $k_1=\sqrt{2}$ and $k_2=1/4$ \cite{Liu}. 
The contrapositive statement on $R^{PC}_\epsilon(h, \nu)\leq1-\gamma$ then gives the necessary condition Eq.~\eqref{eqn: pc_risk_so}.
\end{proof}

\section{Proof of Corollary~\ref{cor:to_thm_pc_risk}} \label{append:proof_cor}
\begin{proof}
We have from Theorem~\ref{thm: pc_risk_so} that the necessary condition for $R^{PC}_\epsilon(h,\nu)\leq1-\gamma$ on $\mathbb{SU}(N)$ is
$\norm{U-V}_2\leq\sqrt{4/N}\lambda_1$ where $\lambda_1=\left[[\ln(2\sqrt{2}/\eta)]^{1/2}+[\ln(2\sqrt{2}/\gamma)]^{1/2}\right]$. Let $\sigma=V\ket{\vec{0}}\bra{\vec{0}}V^{\dagger}$. From the Proof of Theorem 3 in \cite{Liu}, we have
$\norm{U-V}_2^2\geq2N(1-\abs{\braket{\phi|\psi}})$. The Fuchs–van de Graaf inequality for pure states is
\begin{equation}\label{eqn:fuchs-van}
    2-2\sqrt{F(\rho,\sigma)}\leq\norm{\rho-\sigma}_1=2\sqrt{1-F(\rho, \sigma)},
\end{equation}
where the fidelity $F(\rho,\sigma)=\abs{\braket{\phi|\psi}}^2$. Based on Eq.~\eqref{eqn:fuchs-van}, we obtain 
\begin{equation*}
\begin{split}
    2N(1-\abs{\braket{\phi|\psi}})&\geq \frac{2NT(\rho, \sigma)^2}{(1+\abs{\braket{\phi|\psi}})}\geq NT(\rho,\sigma)^2,
\end{split}
\end{equation*}
where $T$ is the trace distance. As such, we need 
\begin{equation*}
\begin{split}
    \sqrt{\frac{4}{N}}\lambda_1 \geq\norm{U-V}_2\geq \sqrt{N}T(\rho,\sigma)=\frac{\sqrt{N}}{2}\norm{\rho-\sigma}_1,
\end{split}
\end{equation*}
which gives $\norm{\rho-\sigma}_1\leq4/N\lambda_1=4d^{-n}\lambda_1$. 
\raggedbottom
\par We translate this upper bound on the distance between two density matrices to that between their encoding vectors $g_1(z)$ and $g_1(z')$. Altogether with the necessary condition and Eq.~\eqref{eqn:fuchs-van}, we have 
\begin{equation}\label{eqn:necessary_vector}
    4d^{-n}\lambda_1\geq \norm{\rho-\sigma}_1 \geq 2-2\sqrt{F(\rho,\sigma)}.
\end{equation}

For density matrices $\rho, \sigma\in \mathcal{X}$ respective to two images, we have $\rho=\ket{\phi}\bra{\phi}=\bigotimes_{i}\ket{\phi_i}\bigotimes_{i}\bra{\phi_i}=\bigotimes_{i}\ket{\phi_i}\bra{\phi_i}=\bigotimes_{i}\rho_i$ and $\sigma=\bigotimes_{i}\ket{\psi_i}\bra{\psi_i}=\bigotimes_{i}\sigma_i$, which are mapped from images $g_1(z)=\vec{s}$ and $g_1(z')=\vec{t}$, respectively. All $i$-indices run from $1$ to $n$. And $\ket{\phi_i}$ and $\ket{\psi_i}$ are featurized from pixels of value $s_i$ and $t_i$, respectively. It can be shown by induction that
\begin{equation} \label{eqn:fidelity}
    F(\rho,\sigma)= \prod_i\cos^{2(d-1)}\left(\abs{s_i-t_i}\frac{\pi}{2}\right).
\end{equation}
For $d=2$, we have that $F(\rho,\sigma)=\tr(\bigotimes_{i}\rho_i\bigotimes_{i}\sigma_i)=\prod_{i} \tr(\rho_i\sigma_i)=\prod_i\abs{\braket{\phi_i|\psi_i}}^2=\prod_i\cos^2(\abs{s_i-t_i}\pi/2)$. It then suffices to show  $\braket{\phi_i|\psi_i}=\cos^{d-1}(\abs{s_i-t_i}\pi/2)$ for the qudit encoding $d>2$. We drop all $\pi/2$ factors and the subscripts $i$ in $s_i$ and $t_i$ hereafter. Suppose for $d=k$, we have $\braket{\phi_i|\psi_i}$ equal to
\begin{equation} \label{eqn:proof_prop_2_1}
\begin{split}
    \sum_{j=1}^k&\binom{k-1}{j-1}\cos^{k-j}(s)\cos^{k-j}(t)\sin^{j-1}(s)\sin^{j-1}(t)\\
    &=\cos^{k-1}(s-t).
\end{split}
\end{equation}
Then for $d=k+1$, we have $\braket{\phi_i|\psi_i}$ equal to
\begin{equation}
\begin{split}
    &\sum_{j=1}^{k+1}\binom{k}{j-1}\cos^{k+1-j}(s)\cos^{k+1-j}(t)\sin^{j-1}(s)\sin^{j-1}(t)\\
    &=\cos(s) \cos(t)\left[\sum_{j=1}^{k}\beta\binom{k}{j-1}\cos^{k-j}(s)\cos^{k-j}(t)\right.\\
    &\quad\ \left. \sin^{j-1}(s)\sin^{j-1}(t)\vphantom{\sum_{j=1}^{k}}\right]+\sin(s)\sin(t)\left[\sum_{j=2}^{k+1}(1-
    \beta)\right.\\
    &\ \ \ \left.\binom{k}{j-1}\cos^{k+1-j}(s)\cos^{k+1-j}(t)\sin^{j-2}(s)\sin^{j-2}(t)\vphantom{\sum_{j=1}^{k}}\right],
\end{split}
\end{equation} 
where $\beta=(k+1-j)/k$.

Identifying the two expressions in the square brackets as both equal to Eq.~\eqref{eqn:proof_prop_2_1}, we obtain the desired outcome $\braket{\phi_i|\psi_i}=\cos^{k}(s-t)$, and the induction completes.

Combining Eq.~\eqref{eqn:necessary_vector} and Eq.~\eqref{eqn:fidelity}, we have
\begin{equation}\label{eqn:append_coro_cosine}
\begin{split}
    4d^{-n}\lambda_1&\geq 2-2\prod_i \cos^{d-1}\left(\abs{s_i-t_i}\frac{\pi}{2}\right)\\ 
    &\geq 2-2\cos^{(d-1)n}\left(\frac{\sum_i\abs{s_i-t_i}}{n}\frac{\pi}{2}\right).
\end{split}
\end{equation}
where the last inequality follows from the inequality $\cos^n(\sum_{i} x_i/n)\geq\prod_{i}\cos(x_i)$. It can be readily shown for $n\geq2$ using the following trick. Consider any pair $x_i$ and $x_j$ and let $x_m$ be their arithmetic average so $x_i=x_m+d$ and $x_j=x_m-d$ for some $d\neq0$. Then $\cos(x_i)\cos(x_j)=\cos(x_m+d)\cos(x_m-d)=\cos^2(x_m)-\sin^2(d)\leq\cos^2(x_m)$. Therefore, one can maximize the overall cosine product, while maintaining the sum of the arguments, by replacing any pair $\cos(x_i)$ and $\cos(x_j)$ with $\cos(x_m)$ and $\cos(x_m)$, and successively replacing every pair till every factor converges to $\cos(\sum_i x_i/n)$ with the same argument.

Solving for $\sum_i\abs{s_i-t_i}=\norm{g_1(z)-g_1(z')}_1$ in Eq.~\eqref{eqn:append_coro_cosine} yields the upper bound on the perturbation size in $(\mathcal{I},\ell^1)$.
\end{proof}

\section{Proof of Proposition~\ref{prop:g}}\label{append:proof_omega_smooth}
\begin{proof}
We decompose $g$ into $g_2\circ g_1$ where $g_1:(\mathcal{Z},\ell^2)\rightarrow(\mathcal{I},\ell^1)$ is desired to be smooth in practice. It can be generalized to $\ell^p$ norm on $\mathcal{I}$ and similar proof follows since the $\ell^p$ norm of any given vector does not grow with $p$. We have $\norm{g_1(z)-g_1(z')}_1\leq\omega_1(\norm{z-z'}_2),\forall z,z'\in\mathcal{Z}$.

We show that it is also smooth for the qudit encoding $g_2: (\mathcal{I},\ell^1)\rightarrow(\mathcal{X},L^1)$ as in Eq.~\eqref{eqn:qudit}. Applying the qudit feature map and similar to that in Appendix~\ref{append:proof_cor}, it can be shown that
\begin{equation}\label{eqn:rho-sigma}
    \norm{\rho-\sigma}_1= 2\sqrt{1-\prod_i\cos^{2(d-1)}\left(\abs{s_i-t_i}\frac{\pi}{2}\right)}.
\end{equation}

Since $\omega(\cdot)$ is used in an upper bound in Theorem~\ref{thm: in_dist_risk}, we need to obtain the scaling of a lower bound to $\omega(\cdot)$. The $\omega(\cdot)$ that forms a tight upper bound in Eq.~\eqref{eqn:modulus_continuity} must have $\omega(\norm{z-z'}_2)$ upper bounding Eq.~\eqref{eqn:rho-sigma} for arbitrary $z,z'\in\mathcal{Z}$. Hence, it is equivalent to find the scaling of a lower bound to Eq.~\eqref{eqn:rho-sigma}. That is, we have $\forall z, z' \in\mathcal{Z}$,  
\begin{equation*}
\begin{split}
    \omega(\norm{z&-z'}_2)\geq2\sqrt{1-\prod_i\cos^{2(d-1)}\left(\abs{s_i-t_i}\frac{\pi}{2}\right)}\\
    &\geq2\sqrt{1-\cos^{2(d-1)n}\left(\frac{\sum_i\abs{s_i-t_i}}{n}\frac{\pi}{2}\right)}\\
    &=2\sqrt{1-\cos^{2(d-1)n}\left(\frac{\pi}{2n}\norm{g_1(z)-g_1(z')}_1\right)},
\end{split}
\end{equation*}
where the second inequality follows from the inequality $\cos^n(\sum_{i} x_i/n)\geq\prod_{i}\cos(x_i)$ proven for Eq.~\eqref{eqn:append_coro_cosine}. Since the above inequality holds for any $z, z'$ such that $\norm{z-z'}_2=\tau$ for any $\tau$, and since we assume $\omega(\cdot)$ forms a tight upper bound in Eq.~\eqref{eqn:modulus_continuity}, $g$ is smooth with
\begin{equation*}
    \omega(\tau)\geq\sqrt{1-\cos^{2n(d-1)}\left(\frac{\pi}{2n}\omega_1(\tau)\right)},\quad \forall\tau>0.
\end{equation*}

In terms of the scaling with respect to $n$ and $d$, if $\omega_1(\cdot)=\Omega(1)$, such as when $g_1$ is Lipschitz continuous, we have $\omega(\cdot)=\Omega(\sqrt{d/n})$.
\end{proof}

\section{Proof of Theorem~\ref{thm: in_dist_risk}}\label{append:proof_major_thm}
\begin{proof}
  If letting $\epsilon_{in}\geq \omega(\sqrt{\ln{[\pi/(2\gamma^2)]}})$, then $\gamma\geq\sqrt{\pi/2}\exp(-\omega^{-1}(\epsilon_{in})^2/2)$. 
  By the definition of the generator and the latent space, we have $\mathcal{N}_m(g^{-1}(\rho))=\xi(\rho),\ \forall\rho\in\mathcal{S}\subseteq\mathcal{X}$.
Let us define $h^i_\rightarrow=\{\rho\in h^i|\mathrm{d}(\rho,\cup_{j\neq i}h^j)\leq\epsilon_{in}\}$ which is the set of density matrices that are at positive distance at most $\epsilon_{in}$ from $\cup_{j\neq i}h^j$, then following Definition~\ref{def:in_dist_risk},
\begin{equation}\label{eqn:eqv_prob}
\begin{split}
     R^{PC}_{\epsilon_{in}}(h, \xi)&=\Pr_{\rho\leftarrow \xi}[\min_{\sigma\in\mathcal{S}}\{\norm{\sigma-\rho}_1|h(\sigma)\neq h(\rho)\}\leq \epsilon_{in}]\\
     &=\xi(\cup_i h^i_\rightarrow))=\mathcal{N}_m(g^{-1}(\cup_i h^i_\rightarrow)),
\end{split}
\end{equation}
  since $h^i_\rightarrow$ are disjoint for different class $i$. Hence, it can be shown that $R^{PC}_{\epsilon_{in}}(h, \xi)\geq1-\gamma$ when $\xi(h^i)\leq1/2,\forall i$ from Theorem 1 in \cite{Fawzi2018}. The contrapositive yields the necessary condition Eq.~\eqref{eqn:gen_eps}. For completeness, we present our condensed version of the proof below.
  
  We write the cumulative distribution function of the standard Gaussian distribution $\mathcal{N}(0,1)$ as $\Phi(x)=1/\sqrt{2\pi}\int_{-\infty}^x \exp(-u^2/2) \mathrm{d} u$. 
  \begin{theorem}[(Gaussian isoperimetric inequality)\cite{borell_brunn-minkowski_1975, ledoux_concentration_2001}]\label{thm_append:gaussian_isoperimetric}
      Let $\mathcal{N}_m$ be the canonical Gaussian measure on $\mathbb{R}^m$. Let $\Sigma\subseteq \mathbb{R}^m$ be any Borel set and let $\Sigma_\epsilon=\{z\in\mathbb{R}^m|\exists z'\in \Sigma\ \text{s.t.}\  \norm{z-z'}_2\leq\epsilon\}$. If $\mathcal{N}_m(\Sigma)=\Phi(a)$ then $\mathcal{N}_m(\Sigma_\epsilon)\geq\Phi(a+\epsilon)$.
  \end{theorem}
  
  \begin{lemma}[\cite{Fawzi2018}]\label{lemma_append:cdf}
   Let $p\in[1/2,1]$, we have for all $\eta>0$,
   \begin{equation}\label{eqn:append_lemma1_1}
   \begin{split}
       \Phi(\Phi^{-1}(p)+\eta)&\geq1-(1-p)\sqrt{\frac{\pi}{2}}e^{-\frac{\eta^2}{2}}.
    \end{split}
   \end{equation}
   If $p=1-1/K$ for $K\geq5$ and $\eta\geq1$, we have
   \begin{equation}\label{eqn:append_lemma1_2}
   \begin{split}
       \Phi(\Phi^{-1}(1-\frac{1}{K})+\eta)&\geq1-\frac{1}{K}\sqrt{\frac{\pi}{2}}e^{-\frac{\eta^2}{2}}e^{-\eta\sqrt{\log\left(\frac{K^2}{4\pi\log(K)}\right)}}.
    \end{split}
   \end{equation}
  
  \end{lemma}
   We first introduce the following sets in the latent space $(\mathbb{R}^m, \ell^2, \mathcal{N}_m)$: $H^i=g^{-1}(h^i)$ and $H^i_\rightarrow=\{z\in H^i|\mathrm{d}(z,\cup_{j\neq i}H^j)\leq\omega^{-1}(\epsilon_{in})\}$. We note that $H^i_\rightarrow\bigcup \cup_{j\neq i}H^j$ is the set of points that are at distance at most $\omega^{-1}(\epsilon_{in})$ from $\cup_{j\neq i}H^j$. Then by Theorem~\ref{thm_append:gaussian_isoperimetric} applied with $\Sigma=\cup_{j\neq i}H^j$ and $a=a_{\neq i}\equiv\Phi^{-1}(\mathcal{N}_m(\cup_{j\neq i}H^j))$, we have $\mathcal{N}_m(H^i_\rightarrow)+\mathcal{N}_m(\cup_{j\neq i}H^j)\geq\Phi(a_{\neq i}+\omega^{-1}(\epsilon_{in}))$. Rearranging, $\mathcal{N}_m(H^i_\rightarrow)\geq \Phi(a_{\neq i}+\omega^{-1}(\epsilon_{in}))-\Phi(a_{\neq i})$. As $H^i_\rightarrow$ are disjoint for different class $i$, we have
  \begin{equation*}
      \mathcal{N}_m(\cup_i H^i_\rightarrow)\geq \sum_{i=1}^K \left[\Phi(a_{\neq i}+\omega^{-1}(\epsilon_{in}))-\Phi(a_{\neq i})\right].
  \end{equation*}
  By the definition of $\omega(\cdot)$, we have $g(H^i_\rightarrow)\subseteq h^i_\rightarrow$. It leads to $\mathcal{N}_m(g^{-1}(h^i_\rightarrow))\geq\mathcal{N}_m(H^i_\rightarrow)$ and $\mathcal{N}_m(\cup_i g^{-1}(h^i_\rightarrow))\geq\mathcal{N}_m(\cup_i H^i_\rightarrow)$. Therefore, we obtain the result for arbitrary decision boundary,
  \begin{equation*}
      \mathcal{N}_m(\cup_i g^{-1}(h^i_\rightarrow)) 
      \geq \sum_{i=1}^K \left[\Phi(a_{\neq i}+\omega^{-1}(\epsilon_{in}))-\Phi(a_{\neq i})\right].
  \end{equation*}
  Equivalently by Eq.~\eqref{eqn:eqv_prob},
  \begin{equation*}
      R^{PC}_{\epsilon_{in}}(h, \xi)
      \geq \sum_{i=1}^K \left[\Phi(a_{\neq i}+\omega^{-1}(\epsilon_{in}))-\Phi(a_{\neq i})\right].
  \end{equation*}
  
  Suppose $\xi(h^i)=\mathcal{N}_m(H^i)\leq1/2$ and $\mathcal{N}_m(\cup_{j\neq i}H^j)\geq 1/2,\forall i$. Using Eq.~\eqref{eqn:append_lemma1_1} in Lemma~\ref{lemma_append:cdf} in the second inequality below,
  \begin{equation*}
    \begin{split}
        R^{PC}_{\epsilon_{in}}&(h,\xi) \geq \sum_{i=1}^K \left[\Phi(\Phi^{-1}(\mathcal{N}_m(\cup_{j\neq i}H^j))+\omega^{-1}(\epsilon_{in}))\right.\\
        &\left.\hspace{5.5em}-\mathcal{N}_m(\cup_{j\neq i}H^j)\right]\\
        &\geq \sum_{i=1}^K\left[1-(1-\mathcal{N}_m(\cup_{j\neq i}H^j))\sqrt{\frac{\pi}{2}}e^{\frac{-\omega^{-1}(\epsilon_{in})^2}{2}}\right.\\
        &\left.\hspace{3.5em}-\mathcal{N}_m(\cup_{j\neq i}H^i)\vphantom{\frac{\pi}{2}}\right]\\
        &=  \left(1-\sqrt{\frac{\pi}{2}}e^{\frac{-\omega^{-1}(\epsilon_{in})^2}{2}}\right)\sum_{i=1}^K\left[1-\mathcal{N}_m(\cup_{j\neq i}H^i)\right]\\
        &=1-\sqrt{\frac{\pi}{2}}e^{\frac{-\omega^{-1}(\epsilon_{in})^2}{2}}> 1-\gamma,
    \end{split}
    \end{equation*}
    provided that $\gamma>\sqrt{\pi/2}\exp(-\omega^{-1}(\epsilon_{in})^2/2)$. The contrapositive yields the results in our Theorem~\ref{thm: in_dist_risk} that $\epsilon_{in}\leq \omega(\sqrt{\ln{[\pi/(2\gamma^2)]}})$ is necessary for $R^{PC}_{\epsilon_{in}}(h, \xi)\leq1-\gamma$.
    
    When there are at least 5 equiprobable classes \cite{Fawzi2018}, substituting Eq.~\eqref{eqn:append_lemma1_2} in Lemma~\ref{lemma_append:cdf} into the above inequality yields
    \begin{equation*}
        R^{PC}_{\epsilon_{in}}(h,\xi) \geq 
        1-\sqrt{\frac{\pi}{2}}e^{\frac{-\omega^{-1}(\epsilon_{in})^2}{2}}e^{-\epsilon_{in}\sqrt{\log\left(\frac{K^2}{4\pi\log(K)}\right)}}.
    \end{equation*}
    Hence, the in-distribution robustness of $h$ decreases with the number of equiprobable classes.
    
     Alternatively, a numerically looser upper bound on $\epsilon_{in}$ can be derived from the fact that $(\mathbb{R}^m,\ell^2,\mathcal{N}_m)$ resembles a normal L\'evy family but the concentration function decays independently of $N$. By Theorem~\ref{thm_append:gaussian_isoperimetric}, any Borel set $\Sigma$ there such that $\mathcal{N}_m(\Sigma)=\Phi(a)$ satisfies $\mathcal{N}_m(\Sigma_\epsilon)\geq \Phi(a+\epsilon)$. In particular, for all Borel sets $A$ with measure at least $1/2$, we have $a\geq0$ and thus, $1-\mathcal{N}_m(A_\epsilon)\leq1-\Phi(\epsilon)\leq \exp(-\epsilon^2/2)$ where the last inequality follows from the Gaussian tail bound. By definition of the concentration function in Eq.~\eqref{eqn:concentration_func}, $\alpha(\epsilon)=\sup_{A}\{1-\mathcal{N}_m(A_\epsilon)\}\leq\exp(-\epsilon^2/2)$.
     
     By substituting the statement and the proof of Theorem~\ref{thm: pc_risk_so} with $k_1=1$ and $k_2=1/\sqrt{2}$ and $N=1$, we have the following. Let $\eta\in[0,1/2]$ be such that $\mathcal{N}_m(H^l)=\xi(h^l)\leq1-\eta,\ \forall l\in\mathcal{L}$. If $\epsilon_{in}\geq\omega(\sqrt{\ln(4/\gamma^2)}+\sqrt{\ln(4/\eta^2)})$, then by acting $\omega^{-1}(\cdot)$, which is a strictly increasing function, on both sides, we obtain $\omega^{-1}(\epsilon_{in})\geq\sqrt{\ln(4/\gamma^2)}+\sqrt{\ln(4/\eta^2)}$. This implies that $R^{PC}_{\omega^{-1}(\epsilon_{in})}(h,\mathcal{N}_m)\geq1-\gamma$. Since $R^{PC}_{\omega^{-1}(\epsilon_{in})}(h,\mathcal{N}_m)\leq R^{PC}_{\epsilon_{in}}(h,\xi)$ (this is equivalent to $g(H^i_\rightarrow)\subseteq h^i_\rightarrow$), it therefore implies $R^{PC}_{\epsilon_{in}}(h,\xi)\geq1-\gamma$. The contrapositive yields, for $R^{PC}_{\epsilon_{in}}(h,\xi)\leq1-\gamma$, it is necessary to have $\epsilon_{in}\leq\omega(\sqrt{\ln(4/\gamma^2)}+\sqrt{\ln(4/\eta^2)})$. When $\eta=1/2$, it can be verified that this necessary upper bound is looser than that in Theorem~\ref{thm: in_dist_risk} for the same $\gamma$.
    
\end{proof}

\section{Proof of Theorem~\ref{thm:sqt}} \label{append:proof_thm_sqt}
\begin{proof}
We have the mapping to obtain a product state density matrix $P:\mathcal{X}\rightarrow\mathcal{X},\ \sigma\mapsto\bigotimes_{i=1}^n \tr_{\{j\neq i\}}\sigma$ where $n$ is the number of qubits. This is not a CPTP map on the set of $d^n\times d^n$ density matrices $\mathcal{X}$ since it is non-linear. Nonetheless, it can be viewed as a CPTP map $\Lambda$ on $\mathcal{X}^{\otimes n}$ as $\Lambda:\mathcal{X}^{\otimes n}\rightarrow \mathcal{X},\ \sigma^{\otimes n}\mapsto \tr_{\{j\neq i\}}([\sigma^{\otimes n}]_i)$ where $[\sigma^{\otimes n}]_i$ denotes the $i$-th copy of $\sigma$, which involves only partial tracing. In particular, for a product state $\rho^{\otimes a}$ with the integer $a\geq1$, $\Lambda(\rho^{\otimes a})=\rho$.

Consider $\rho\in\mathcal{S}\subseteq \mathcal{X}$ an $n$-qubit density matrix, namely $\rho=g(z)$ for some $z\in\mathcal{Z}$. Let $\sigma\in\mathcal{X}$. We have
\begin{equation*}
\begin{split}
    \norm{\rho-P(\sigma)}_1&=\norm{\Lambda(\rho^{\otimes n})-\Lambda(\sigma^{\otimes n})}_1\leq\norm{\rho^{\otimes n}-\sigma^{\otimes n}}_1\\
    &\leq 2\sqrt{1-F(\rho^{\otimes n},\sigma^{\otimes n})}=2\sqrt{1-F(\rho,\sigma)^n},
\end{split}
\end{equation*}
where the first inequality follows from the contractive property of the trace norm under any CPTP map and the last equality follows from the multiplicativity of fidelity with respect to tensor products. By Eq.~\eqref{eqn:fuchs-van}, we have $F(\rho,\sigma)\geq(1-\norm{\rho-\sigma}_1/2)^2$. Substituting in, we obtain
\begin{equation*}
    \norm{\rho-P(\sigma)}_1\leq2\sqrt{1-\left(1-\frac{\norm{\rho-\sigma}_1}{2}\right)^{2n}}.
\end{equation*}
Let $\tilde{\sigma}\in\mathcal{S}$ be the closest in-distribution sample to $P(\sigma)$, which can be found by fitting parameters $\{s_i\}$ in Eq.~\eqref{eqn: product_states}. Therefore, $\norm{P(\sigma)-\tilde{\sigma}}_1\leq\norm{P(\sigma)-\rho}$. We then obtain
\begin{equation}\label{eqn:proof_sqt_1}
\begin{split}
    \norm{\rho-\tilde{\sigma}}_1&\leq\norm{\rho-P(\sigma)}_1+\norm{P(\sigma)-\tilde{\sigma}}_1\\
    &\leq 4\sqrt{1-\left(1-\frac{\norm{\rho-\sigma}_1}{2}\right)^{2n}}.
\end{split}
\end{equation}
Recall that for the quantum classifier $\tilde{h}$, $\tilde{h}(\sigma)=h(\tilde{\sigma})$. Taking minimum over all $\sigma$ such that $\tilde{h}(\sigma)\neq \tilde{h}(\rho)$ (i.e., $h(\tilde{\sigma})\neq h(\rho)$),
\begin{equation}\label{eqn:proof_sqt_2}
\begin{split}
\varepsilon_{in}(\rho)&\leq\min \{\norm{\rho-\tilde{\sigma}}_1\}\\
&\leq 4\sqrt{1-\left(1-\frac{\min\{\norm{\rho-\sigma}_1\}}{2}\right)^{2n}},
\end{split}
\end{equation}
we obtain
\begin{equation} \label{eqn:proof_sqt_3}
\varepsilon_{in}(\rho)\leq 4\sqrt{1-\left(1-\frac{\varepsilon_{unc}(\rho)}{2}\right)^{2n}}.
\end{equation}

Notice that to obtain an inequality between $\varepsilon_{in}(\rho)$ and $\varepsilon_{unc}(\rho)$ like in Eq.~\eqref{eqn:proof_sqt_3}, it is sufficient to have Eq.~\eqref{eqn:proof_sqt_2} hold after taking the minimum, and it is not necessary to have Eq.~\eqref{eqn:proof_sqt_1} hold for a generic $\sigma$. Since for $n$-qubit density matrices which are separable with respect to some equal bipartition of the system, denoted as $\{\rho_b\}$, form a dense subset \cite{orus_weakly-entangled_2004}, we can effectively realize the same minimum in Eq.~\eqref{eqn:proof_sqt_2} over $\sigma\in\{\rho_b\}$ such that $\tilde{h}(\sigma)\neq\tilde{h}(\rho)$ instead. For equal bipartite states, the number of copies to make a CPTP map $\Lambda'$ acting on them to obtain $P(\sigma)$ reduces to $n/2$ if $n$ is even and reduces to $(n+1)/2$ if $n$ is odd. For instance, given a 4-qubit $\sigma$ whose qubit $1$ is only entangled with $2$ and qubit $3$ is only entangled with $4$, $\Lambda'(\sigma^{\otimes2})=\tr_{\{1,3\}}(\sigma)\otimes \tr_{\{2,4\}}(\sigma)=P(\sigma)=\Lambda(\rho^{\otimes4})$. Therefore, we can replace the exponent $1/(2n)$ in Eq.~\eqref{eqn:proof_sqt_3} with $1/n$ for even $n$ and $1/(n+1)$ for odd $n$.

We recall $\varepsilon_{unc}(\rho)\leq \varepsilon_{in}(\rho),\ \forall \rho\in\mathcal{X}$ and rearrange,
\begin{equation*}
    2-2\left(1-\frac{\varepsilon_{in}(\rho)^2}{16}\right)^{\frac{1}{n_e}}\leq \varepsilon_{unc}(\rho)\leq \varepsilon_{in}(\rho),
\end{equation*}
where $n_e=n$ for even $n$ and $n_e=n+1$ for odd $n$.
\end{proof}

\end{appendix}
\end{document}